\documentclass[aps,a4paper,twocolumn,superscriptaddress,eqsecnum,longbibliography,floatfix]{revtex4-1}

\usepackage[T1]{fontenc}
\usepackage{newtxtext}
\usepackage{newtxmath}

\usepackage{amsmath}
\usepackage{mathtools}
\usepackage{amsfonts}
\usepackage{bm}

\usepackage{booktabs}
\usepackage{makecell}

\usepackage{enumitem}

\usepackage{graphicx}

\usepackage{color}
\usepackage{xcolor}
\usepackage[colorlinks,
            citecolor=blue,
            linkcolor=red,
            urlcolor=blue]{hyperref}

\begin{document}

\title{Quantum magnetism on small-world networks}

\author{Maxime Dupont}
\affiliation{Department of Physics, University of California, Berkeley, California 94720, USA}
\affiliation{Materials Sciences Division, Lawrence Berkeley National Laboratory, Berkeley, California 94720, USA}

\author{Nicolas Laflorencie}
\affiliation{Laboratoire de Physique Th\'eorique, IRSAMC, Universit\'e de Toulouse, CNRS, UPS, 31062 Toulouse, France}

\begin{abstract}
    While classical spin systems in random networks have been intensively studied, much less is known about quantum magnets in random graphs. Here, we investigate interacting quantum spins on small-world networks, building on mean-field theory and extensive quantum Monte Carlo simulations. Starting from one-dimensional (1D) rings, we consider two situations: all-to-all interacting and long-range interactions randomly added. The effective infinite dimension of the lattice leads to a magnetic ordering at finite temperature $T_\mathrm{c}$ with mean-field criticality. Nevertheless, in contrast to the classical case, we find two distinct power-law behaviors for $T_\mathrm{c}$ versus the average strength of the extra couplings. This is controlled by a competition between a characteristic length scale of the random graph and the thermal correlation length of the underlying 1D system, thus challenging mean-field theories. We also investigate the fate of a gapped 1D spin chain against the small-world effect.
\end{abstract}

\maketitle

\section{Introduction}

\subsection{Complex networks and the small-world effect}

Understanding complex networks is at the heart of many scientific fields~\cite{scott_social_1988,watts_collective_1998,barabasi_emergence_1999,barrat_properties_2000,strogatz_exploring_2001,girvan_community_2002,albert_statistical_2002,barrat_architecture_2004,dorogovtsev_critical_2008,arenas_synchronization_2008,pastor-satorras_epidemic_2015}, such as computer science, mathematics, physics, biology, sociology, epidemiology, etc. During the past two decades, critical phenomena arising in such random topologies have emerged as a key subject of intense research in statistical physics~\cite{albert_statistical_2002,dorogovtsev_critical_2008}.

A complex network is a graph with nontrivial and random properties, as opposed to periodic (or quasiperiodic) lattices of finite dimension. There are two main features which contrast with regular graphs: (i) a fluctuating connectivity (a certain proportion of the links are randomly placed) and (ii) the so-called small-world (SW) effect~\cite{Porter:2012}, which can dramatically   shorten the distances across the network. More precisely, for a finite graph of $N$ sites, the average distance $\overline\ell$ between two arbitrary points, also called the graph diameter, grows slower than any power-law with $N$: $\overline\ell\sim\ln N$, resulting in an infinite effective dimension.

\begin{figure}[!t]
    \includegraphics[angle=0,clip=true,width=\columnwidth]{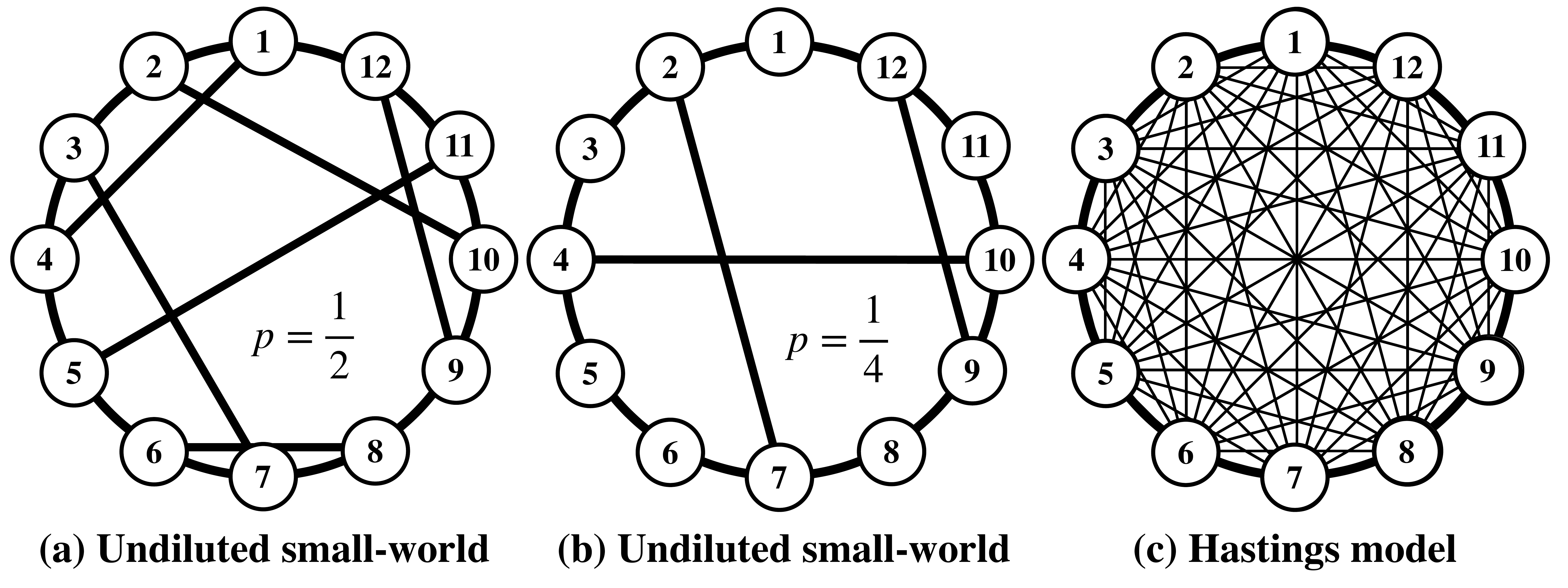}
    \caption{Two types of small-world networks with $N=12$ sites. (a, b) Long-range bonds are added with probability $p$, without diluting the underlying $1$D structure. (c) All possible long-range links are added, but with a reduced strength $\propto 1/N$ vanishing at large sizes.}
    \label{fig:lattices}
\end{figure}

The SW effect occurs in a large class of complex networks, such as  Erd\"os-R\'enyi random graphs~\cite{Erdos1959}, scale-free~\cite{barabasi_emergence_1999} and SW networks~\cite{watts_collective_1998}. For the later case, the most popular SW system is the Watts-Strogatz model~\cite{watts_collective_1998} in which one randomly rewire with a probability $p$ each edge of an initial one-dimensional (1D) ring. Shortly after, a variant was proposed in Refs.~\cite{monasson_diffusion_1999,newman_renormalization_1999} by simply adding long-range bonds with  probability $p$, without diluting the underlying 1D structure [see Figs.~\ref{fig:lattices}(a) and~\ref{fig:lattices}(b)]. This undiluted version of the SW network, more amenable to analytical treatments, was argued~\cite{barrat_properties_2000} to bring similar physics as compared to the original SW proposal of Watts and Strogatz. Another simplification was later proposed by Hastings in Ref.~\cite{Hastings2003} with a mean-field (MF) version [see Fig.~\ref{fig:lattices}(c)], where all possible long-range links are added, but with a reduced strength $\propto 1/N$ vanishing at large sizes. This MF variant was introduced to avoid randomness and thus facilitate analytical calculations.

\subsection{Classical magnetism and small-world effect}

A strong consequence of the SW effect is that for any finite concentration $p>0$ of extra long-range links added across a lattice of finite dimension $d$ (exemplified in Fig.~\ref{fig:lattices} for $d=1$), the system will behave as infinite-dimensional $d=\infty$, provided the number of sites, $N$, is large enough, typically exceeding a crossover size $N^\star\sim 1/p$~\cite{barthelemy_small-world_1999,barthelemy_erratum_1999,barrat_properties_2000}. This drastic change in the effective dimension of the problem has attracted a lot attention in the context of interacting classical spin systems~\cite{barrat_properties_2000,gitterman_small-world_2000,kim_xy_2001,hong_comment_2002,herrero_ising_2002,dorogovtsev_ising_2002,igloi_first_2002,goltsev_critical_2003,herrero_ising_2004,lopes_exact_2004,dorogovtsev_potts_2004}, while much less is known for the quantum case~\cite{yi_effect_2003,yi_quantum_2010,baek_quantum_2011,yi_quantum_2015}.

Classical $O(n)$ models on SW networks have been heavily investigated for $n=1$ (Ising)~\cite{barrat_properties_2000,gitterman_small-world_2000,hong_comment_2002,herrero_ising_2002,lopes_exact_2004}, and to a lesser extent for $n=2$ (XY)~\cite{kim_xy_2001,medvedyeva_dynamic_2003}. In both cases, MF theory (expected above $d_\mathrm{u}=4$) was found to describe the critical properties. Note, however, that scale-free networks with power-law distributed connectivities~\cite{barabasi_emergence_1999} do not necessarily display MF behavior, depending on the power-law exponent of the connectivity distribution~\cite{dorogovtsev_ising_2002,goltsev_critical_2003,herrero_ising_2004,de_nigris_critical_2013}. nonuniversal and non-MF behaviors have also been reported in SW networks where the long-range interactions~\cite{jeong_phase_2003} or the branching probability~\cite{chatterjee_phase_2006} decay as a power law with the distance. To some extent, this is reminiscent of early renormalization-group results for $n$-vector models with power-law decaying interactions~\cite{fisher_critical_1972,brezin_critical_1976}.

\subsection{Small-world quantum magnets}

In this work, we want to address the following question: Is there some specificity of quantum spins as compared to the aforementioned classical results? To this end, we will focus on spin-$1/2$ quantum magnets on the SW  geometries depicted in Figs.~\ref{fig:lattices}(b) and~\ref{fig:lattices}(c). The Hamiltonian is made of two components, a short-range part $\mathcal{H}_\mathrm{1D}$ and a random long-range contribution $\mathcal{H}_\mathrm{LR}$. For the short-range piece, we choose the XXZ Hamiltonian, defined on a ring by
\begin{eqnarray}
    {\cal{H}}_\mathrm{1D}&=&J\sum\nolimits_{i=1}^{N}h_{i,i+1}^\Delta,\nonumber\\
    {\rm{with}}\quad h_{i,i+1}^\Delta&=&S_i^xS_{i+1}^x+S_i^y S_{i+1}^y+\Delta S_i^z S_{i+1}^z,
    \label{eq:H1D}
\end{eqnarray}
with periodic boundary conditions. $\Delta$ is the Ising anisotropy parameter, and the long-range part, which describes interactions beyond nearest neighbors, takes a similar XXZ form,
\begin{equation}
    {\cal{H}}_\mathrm{LR}=\sum\nolimits_{i,j}J_{ij}^\mathrm{LR}\,h_{i,j}^\Delta,\quad|i-j|>1.
    \label{eq:HLR}
\end{equation}
In the rest of the paper, we will focus on two emblematic cases:

\begin{enumerate}[label=\arabic*),leftmargin=1\parindent]
    \item The ferromagnetic XY model with $\Delta=0$, and all couplings negative: $J<0$ and $J_{ij}^\mathrm{LR}=-|J_{ij}^\mathrm{LR}|$.
    \item The (staggered) antiferromagnetic Heisenberg model defined by $\Delta=1$, $J>0$, and staggered couplings beyond nearest-neighbor $J_{ij}^\mathrm{LR}=-(-1)^{|i-j|}|J_{ij}^\mathrm{LR}|$ which prevent magnetic frustration. This alternating exchange is well known to enhance antiferromagnetic correlations~\cite{PhysRevB.69.144412,Laflorencie_2005}.
\end{enumerate}

\subsubsection{Undiluted small-world}

Starting with an $N$-site ring, the undiluted SW model, Fig.~\ref{fig:lattices}(b), is controlled by a branching parameter $0<p\le0.5$ such that we randomly draw $\left\lfloor p N \right\rfloor$ long-ranged links $(i,j)$ having a coupling strength $|J_{ij}^\mathrm{LR}|=J'$, while $J_{ij}^\mathrm{LR}=0$ for all other pairs. The average connectivity is therefore ${\overline{z}}=2+2p$, and the average strength of extra long-range couplings is,
\begin{equation}
    \overline{J'(p)}=2pJ'.
    \label{eq:Jpi}
\end{equation}

\subsubsection{Hastings model}

As shown in Fig.~\ref{fig:lattices}(c), the MF version of the SW networks~\cite{Hastings2003} is built by distributing the long-range couplings over all sites with  $|J_{ij}^\mathrm{LR}|=2pJ'/N$ for all pairs $|i-j|>1$. This model has no randomness, and the extra-couplings have a total strength
\begin{equation}
    {\overline{J'(p)}}=2pJ'\frac{N-3}{N}~\longrightarrow~2pJ'\quad (N\to+\infty),\label{eq:Jpii}
\end{equation}
thus making this model equivalent to the undiluted small-world from an energetic point of view, while the connectivity of the Hastings model is extensive $z=N$.

\subsection{Structure of the paper}

The rest of the paper is organized as follows. In Sec.~\ref{sec:MF}, we review previous results on classical spin systems and discuss two MF theories for SW networks. Because of the effective infinite dimensionality of the lattice, one expects a temperature phase transition on this geometry. Interestingly, the two approaches lead to different qualitative behaviors of the critical temperature $T_\mathrm{c}$ with the average strength of extra long-range couplings $\overline{J'(p)}$, as defined in Eqs.~\eqref{eq:Jpi} and~\eqref{eq:Jpii}. The first method is based on a comparison between the thermal correlation length of the system without the extra couplings and a purely geometric quantity: the average distance between two shortcuts. The other approach is based on the random phase approximation (RPA). In Sec.~\ref{sec:1dmf}, we consider these approximate MF treatments for SW graphs built on top of $1$D quantum spin chains for the classical Ising chain and the quantum $S=1/2$ XXZ chain model. In Sec.~\ref{sec:qmc}, we then treat SW systems exactly with quantum Monte Carlo (QMC) simulations that we compare to the MF approaches. We find that while the physics of the Hastings model is exactly captured by the RPA, the undiluted SW system may experience a crossover from one MF behavior to the other as a function of the branching parameter $p$. In order to go beyond gapless XXZ physics, we also explore the fate of a gapped 1D dimerized chain against the SW effect. To conclude, we present a summary of our findings and discuss a few perspectives of our study in Sec.~\ref{sec:conclusion}.

\section{Mean-field theory}
\label{sec:MF}

\subsection{Ising and XY models: discussion of previous results}

As expected from the infinite-dimensional nature of SW networks, several authors agreed on the MF nature of the finite temperature ordering transition for both classical Ising~\cite{barrat_properties_2000,gitterman_small-world_2000,hong_comment_2002,herrero_ising_2002,lopes_exact_2004} and XY~\cite{kim_xy_2001} models. In the limit of small branching probability $p\ll 1$, a simple MF argument predicts a critical temperature when the correlation length of the underlying $d$-dimensional lattice $\xi(T)\sim\left|T-T_\mathrm{c}(0)\right|^{-\nu}$ [with  $T_\mathrm{c}(0)$ the $p=0$ critical temperature and $\nu$ the associated critical exponent] becomes of the order of the average distance between two shortcuts, $\zeta_p\sim p^{-1/d}$. This simple argument gives
\begin{equation}
    T_\mathrm{c}^\mathrm{MF}(p)-T_\mathrm{c}(0)\propto J p^{1/d\nu}.
    \label{eq:TcMF}
\end{equation}
For the $d=1$ Ising model where $T_\mathrm{c}(0)=0$ and $\nu= \infty$ since $\xi_0(T)\sim \exp(2J/T)$, the above MF argument yields,
\begin{equation}
    T_\mathrm{c,Ising}^\mathrm{MF}\propto {2J}\bigr/{\ln\bigl(1/p\bigr)},
    \label{eq:1dising}
\end{equation}
in good agreement with the literature~\cite{barrat_properties_2000,gitterman_small-world_2000,lopes_exact_2004}.

However, when the very same  MF reasoning is applied to the classical XY chain, for which $\xi(T)\sim J/T$ at low temperature, we get $T_\mathrm{c,XY}^\mathrm{MF}\propto Jp$, a result in disagreement with Monte Carlo simulations where a surprising $a\ln p +b$ (with $a,b\in\mathbb{R}$) scaling has been found~\cite{kim_xy_2001}. The MF prediction for $T_\mathrm{c}^\mathrm{MF}$ in Eq.~\eqref{eq:TcMF} has been critically analyzed by Hastings in Ref.~\cite{Hastings2003} where a different scaling with the branching probability was found,
\begin{equation}
    {\tilde{T}}_\mathrm{c}^\mathrm{MF}(p)-T_\mathrm{c}(p=0)\propto J p^{1/\gamma},\label{eq:TcRPA}
\end{equation}
with $\gamma$ the critical exponent controlling the susceptibility (associated to the order parameter) of the underlying $d$-dimensional model: $\chi(T)\sim|T-T_\mathrm{c}(0)|^{-\gamma}$ when $T\to T_\mathrm{c}(0)^{+}$. As we will discuss in more detail below, the expression of Eq.~\eqref{eq:TcRPA} is a direct consequence of a random phase approximation treatment of the problem.

When comparing Eq.~\eqref{eq:TcMF} and Eq.~\eqref{eq:TcRPA} with numerical results obtained for the Ising model  by Herrero~\cite{herrero_ising_2002}, Hastings argued in favor of Eq.~\eqref{eq:TcRPA} since $p^{1/\gamma}<p^{{1}/{d\nu}}$ in the $p\to 0$ limit. However, this statement requires that $\gamma<d\nu$ or, equivalently, using Fisher's identity $\gamma=(2-\eta)\nu$~\cite{kardar_2007},
\begin{equation}
    d+\eta>2,
    \label{eq:condition}
\end{equation}
with $\eta$ the anomalous dimension. This condition is fulfilled for classical phase transitions in spin systems, but as we will see below, low-dimensional quantum magnets provide a unique example where the critical temperature $T_\mathrm{c}(p)$ can cross over from Eq.~\eqref{eq:TcRPA} to Eq.~\eqref{eq:TcMF} when $p\to 0$.

\subsection{Random phase approximation}

The random phase approximation~\cite{Scalapino1975,Schulz1996} gives a self-consistent MF estimate for the ordering transition temperature of weakly coupled $d$-dimensional systems, using
\begin{equation}
    T_\mathrm{c}^\mathrm{RPA}=\chi^{-1}_{d}\left(\frac{1}{J_\perp}\right),
    \label{eq:Tcrpa}
\end{equation}
where $\chi^{-1}_{d}$ is the inverse-susceptibility function of the underlying $d$-dimensional system, and $J_\perp$ is the (weak) MF coupling between the $d$-dimensional units (see App.~\ref{app:RPA}).

The RPA expression for the critical temperature of Eq.~\eqref{eq:Tcrpa} has proven to be very useful in the context of weakly coupled low-dimensional systems~\cite{Yasuda2005} such as coupled spin chains and ladders~\cite{thielemann_field-controlled_2009,bouillot_statics_2011,blinder_nuclear_2017}, or layered magnets~\cite{yao_universal_2007,lancaster_magnetic_2007,goddard_experimentally_2008,juhasz_junger_thermodynamics_2009,johnston_magnetic_2011,gibertini_magnetic_2019}. Interestingly, a direct quantitative comparison between exact QMC simulations for various $d=3$ anisotropic spin models and the RPA expression of Eq.~\eqref{eq:Tcrpa} gives~\cite{Yasuda2005,blinder_nuclear_2017,bollmark_dimensional_2020} a very good agreement, but at the expense of reducing the weak coupling $J_\perp$ by a nonuniversal factor $J_\perp\to \alpha J_\perp$ with $\alpha\simeq 0.7$~\cite{PhysRevB.61.6757,Yasuda2005,PhysRevB.74.184407,thielemann_field-controlled_2009,bollmark_dimensional_2020}.

In our SW networks, the long-range branching across the original $d$-dimensional systems induces  extra-couplings of average strength ${\overline{J'(p)}}$, as given by Eqs.~\eqref{eq:Jpi} and \eqref{eq:Jpii}. Using the susceptibility divergence of the bare system at $p=0$,
\begin{equation}
    J\chi(T)\propto\left(\frac{T-T_\mathrm{c}(0)}{J}\right)^{-\gamma},
\end{equation}
the above RPA formula of Eq.~\eqref{eq:Tcrpa} yields
\begin{equation}
    T_\mathrm{c}^\mathrm{RPA}(p)-T_\mathrm{c}(0)\propto J\left(2p\frac{J'}{J}\right)^{1/\gamma},
    \label{eq:TCRPA}
\end{equation}
which recovers Hastings's expression~\cite{Hastings2003}, as given above in Eq.~\eqref{eq:TcRPA}. Here, we notice that the RPA estimate explicitly depends on the shortcut coupling strength $J'$, while the simpler MF expression of Eq.~\eqref{eq:TcMF} does not.

In the absence of finite temperature transition $T_\mathrm{c}(0)=0$ (e.g., for $d=1$, or $d=2$ with continuous symmetry, such as the Heisenberg or XY models), the RPA expression of Eq.~\eqref{eq:TCRPA} is still valid, as we discuss now.

\section{The special case of $d=1$}
\label{sec:1dmf}

\subsection{Ising chain}

We start with a brief discussion of the $d=1$ Ising model. As seen above, a simple MF argument, valid  for the highly diluted limit $p\ll 1$, yields $\xi(T_\mathrm{c})\sim e^{2J/T_\mathrm{c}(p)}\sim 1/p$, which leads to the well-know form of Eq.~\eqref{eq:1dising}~\cite{barrat_properties_2000,gitterman_small-world_2000,lopes_exact_2004}. However, one can also invoke an RPA treatment of this problem, using the exponential divergence of the susceptibility
\begin{equation}
    \chi{^\mathrm{1d\,Ising}}=\frac{1}{T}\exp\left(\frac{2J}{T}\right),
\end{equation}
which gives in the limit $p\ll 1$
\begin{equation}
    T_\mathrm{c}^\mathrm{RPA}=\frac{2J}{\ln\left(\frac{T_\mathrm{c}^\mathrm{RPA}}{2pJ'}\right)}\approx
    \frac{2J}{\ln \left(\frac{J}{pJ'}\right)}.\label{eq:T1DISINGRPA}
\end{equation}
One sees that if shortcut and nearest-neighbor couplings have equal strengths $J=J'$, the RPA of Eq.~\eqref{eq:T1DISINGRPA} becomes equivalent to the simple MF expression of Eq.~\eqref{eq:1dising}.

If $J'<J$, the ordering will be controlled by $T_\mathrm{c}^\mathrm{RPA}<T_\mathrm{c}^\mathrm{MF}$. In the opposite case $J'>J$, the MF temperature $T_\mathrm{c}^\mathrm{MF}< T_\mathrm{c}^\mathrm{RPA}$ will take over because the 1D correlation length at $T_\mathrm{c}^\mathrm{RPA}$ has not reached the  average distance between two shortcuts $\zeta_p\sim 1/p$, and one would  need to further cool down the system to reach this threshold. We therefore expect from this simple example that the transition temperature will be given by the minimum of the two estimates:
\begin{equation}
    T_\mathrm{c}=\min\left(T_\mathrm{c}^\mathrm{RPA},\,T_\mathrm{c}^\mathrm{MF}\right).
\end{equation}

\subsection{XXZ chain: the case of Tomonaga-Luttinger liquids}

\subsubsection{Analytical results}

The spin-$1/2$ XXZ chain model, described by $\mathcal{H}_\mathrm{1D}$ in Eq.~\eqref{eq:H1D}, is a well-known example of a Tomonaga-Luttinger liquid (TLL) in the regime $-1<\Delta\le1$. Among the vast amount of knowledge available for this class of systems~\cite{giamarchi2003quantum}, let us briefly summarize a few of them, in particular the ones useful in the context of an RPA treatment of $d=1$ XXZ SW models. Only two parameters are sufficient to describe the low-energy properties of ${\cal{H}}_\mathrm{1D}$: the velocity of excitations, $u$, and the so-called Luttinger exponent $K$. Their dependence on the Ising anisotropy $\Delta$ are well known~\cite{korepin_bogoliubov_izergin_1993}:
\begin{equation}
    u=\pi\frac{\sqrt{1-\Delta^2}}{2\arccos\Delta},\quad K=\frac{\pi}{2\arccos\left(-\Delta\right)}.
\end{equation}
In the easy-plane regime $\left|\Delta\right|<1$, the dominant correlations are transverse with respect to the Ising anisotropy and power-law decaying at $T=0$~\cite{giamarchi2003quantum},
\begin{equation}
    \bigl\langle S^x_m S^x_n\bigr\rangle =\frac{A_{xx}}{ \left|m-n\right|^{\frac{1}{2K}}}\,{\rm{e}}^{-iq|m-n|}+\cdots,
    \label{eq:SxSx}
\end{equation}
with $q=0$ ($q=\pi$) for ferromagnetic (antiferromagnetic) interactions. The amplitude $A_{xx}$ in Eq.~\eqref{eq:SxSx} is also known exactly~\cite{LUKYANOV1997571}. This quasi-long-range (algebraic) order does not survive at finite temperature where all correlations decay exponentially with a finite correlation length, diverging at low temperature,
\begin{equation}
    \xi(T)\propto uJ\bigr/ T^\nu~~\mathrm{with}~\nu=1.
    \label{eq:xiT}
\end{equation}
In the regime $|\Delta|<1$, the transverse susceptibility, associated to the dominant correlation of Eq.~\eqref{eq:SxSx}, has the following low-$T$ behavior~\cite{giamarchi2003quantum,bouillot_statics_2011}:
\begin{equation}
    \chi_{xx}\bigl(T\bigr) = \frac{A_{xx}\sin\left(\frac{\pi}{4K}\right)B^2\left(\frac{1}{8K}, 1-\frac{1}{4K}\right)}{uJ}\left(\frac{2\pi T}{uJ}\right)^{-2+\frac{1}{2K}},
    \label{eq:1dchi_tll}
\end{equation}
with $B(x,y)=\Gamma(x)\Gamma(y)/\Gamma(x+y)$, making Eq.~\eqref{eq:1dchi_tll} a parameter-free expression.

\subsubsection{Consequences for the critical temperature}
\label{sec:crossover}

\begin{figure}[t!]
    \includegraphics[angle=0,clip=true,width=\columnwidth]{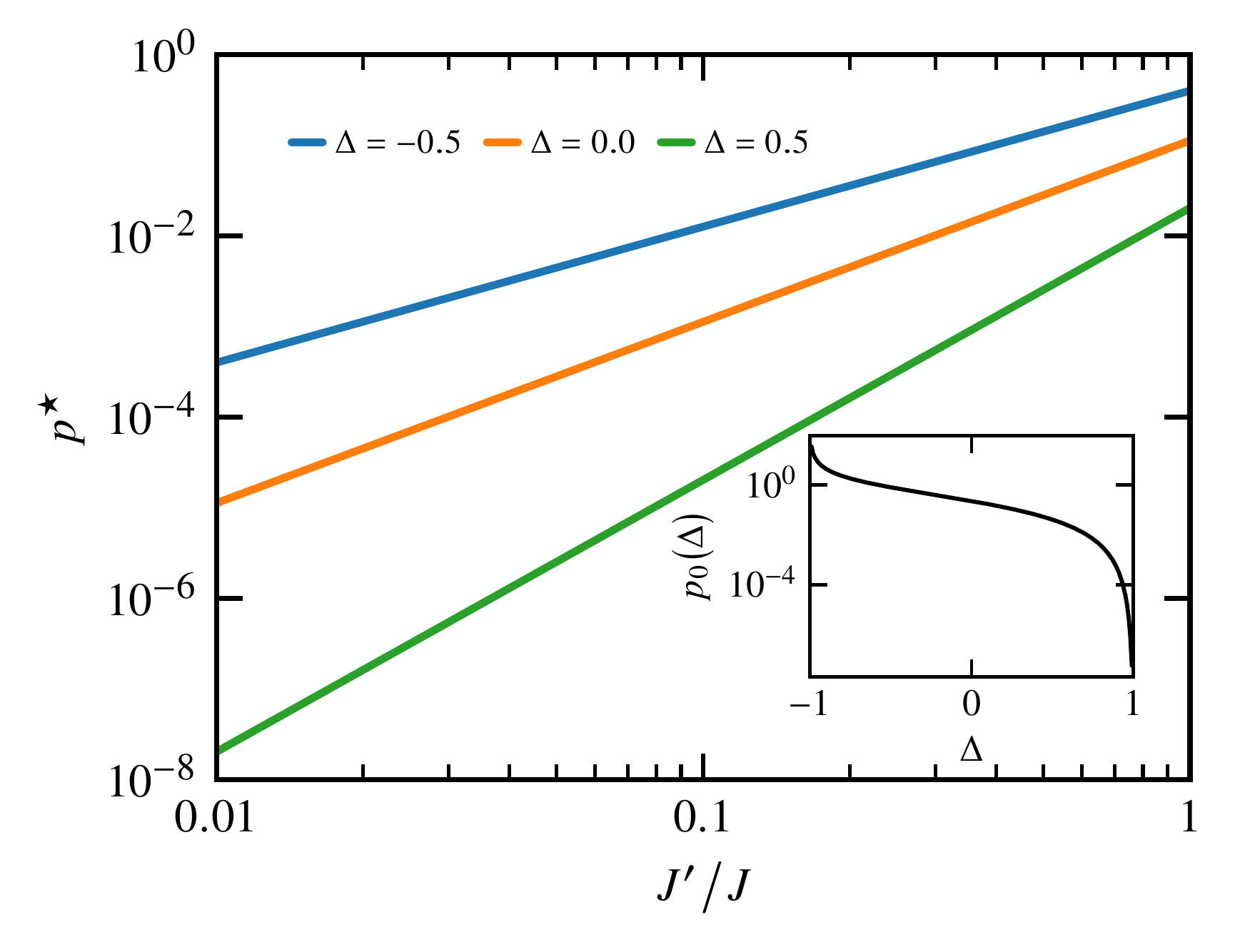}
    \caption{The crossover probability $p^\star$, defined in  Eq.~\eqref{eq:pstar}, is plotted versus the long-range coupling $J'/J$ for different values of Ising anisotropy $\Delta$. The inset shows the behavior of the $J'/J$-independent prefactor of $p^\star$ versus $\Delta$. Note its singular behavior as $|\Delta|\to 1$.}
    \label{fig:crossover_p}
\end{figure}

From the above expression of the transverse susceptibility [Eq.~\eqref{eq:1dchi_tll}], one can identify the susceptibility exponent to be $\gamma=2-\frac{1}{2K}$. Quite interestingly, we see that the above condition Eq.~\eqref{eq:condition} is not fulfilled for TLL with $K=(2\eta)^{-1}>1/2$, which applies to the entire easy-axis regime ($-1\le \Delta <1$), except at the isotropic point. Inverting Eq.~\eqref{eq:1dchi_tll} yields a parameter-free expression for the RPA estimate of the critical temperature,
\begin{equation}
    T_\mathrm{c}^\mathrm{RPA}(p)=uJf\bigl(K,A_{xx}\bigr) \left(\frac{2pJ'}{uJ}\right)^\frac{2K}{4K-1},
    \label{eq:1dTC_tll}
\end{equation}
with the dimensionless prefactor
\begin{equation}
    f\bigl(K,A_{xx}\bigr)=\frac{1}{2\pi}\left[A_{xx} \sin\left(\frac{\pi}{4K}\right)B^2\left(\frac{1}{8K}, 1-\frac{1}{4K}\right)\right]^\frac{2K}{4K-1}.
\end{equation}
When comparing the RPA prediction with the simple MF expression of Eq.~\eqref{eq:TcMF} using the temperature dependence of the correlation length of Eq.~\eqref{eq:xiT},
\begin{equation}
    T_\mathrm{c}^\mathrm{MF}(p)=2uJp,
    \label{eq:1DMF}
\end{equation}
we anticipate a crossover at low branching probability $p^\star$ from an RPA regime of Eq.~\eqref{eq:1dTC_tll} to the linear MF regime of Eq.~\eqref{eq:1DMF} (provided that $K>1/2$). This occurs when $T_\mathrm{c}^\mathrm{RPA}(p^\star)=T_\mathrm{c}^\mathrm{MF}(p^\star)$, meaning that
\begin{equation}
    p^\star=p_0(\Delta)\left({\frac{J'}{J}}\right)^{\mu_\Delta},
    \label{eq:pstar}
\end{equation}
where $p_0(\Delta)=u^{-\mu_\Delta} \,\left[f(K,A_{xx})\right]^{1+\mu_\Delta}$ is plotted in Fig.~\ref{fig:crossover_p}(inset) as a function of the Ising anisotropy, and the exponent $\mu_\Delta=\pi/\arccos(\Delta)$ varies between $1$ for $\Delta=-1$ and $+\infty$ when $\Delta\to 1$. Equation~\eqref{eq:pstar} is plotted against $J'/J$ in Fig.~\ref{fig:crossover_p} for various anisotropies $\Delta$. This defines the range of validity of the RPA expression for the critical temperature, Eq.~\eqref{eq:1dTC_tll}, for $p>p^\star$. Below $p^\star$, the simpler linear MF argument, Eq.~\eqref{eq:1DMF}, is expected.

The antiferromagnetic Heisenberg case ($\Delta=1$) is more subtle since the TLL parameter $K=1/2$ and logarithmic corrections~\cite{Affleck_1989,PhysRevB.43.8217,PhysRevLett.73.332,PhysRevB.56.13681,Affleck_1998,PhysRevB.63.140412,Barzykin_2000,PhysRevB.94.144409} are expected in the temperature dependence of both the correlation length and the staggered susceptibility. This will be discussed in more detail in the following (Sec.~\ref{sec:s12_af_chi}).

\subsection{Quantum Monte Carlo results for the $d=1$ susceptibilities}
\label{sec:susc_qmc}

We simulate the $S=1/2$ XXZ chain model, Eq.~\eqref{eq:H1D}, at finite temperature $T$ with QMC, using the stochastic series expansion with directed loop updates~\cite{syljuaasen2002,sandvik2010,sandvik2019}.

Noting $h_\mathrm{sb}$ is a symmetry-breaking field coupled to the order parameter $\langle m\rangle$, the linear response function (susceptibility $\chi$) takes the form
\begin{equation}
    \chi = \frac{\partial\bigl\langle m\bigl(h_\mathrm{sb}\bigr)\bigr\rangle}{\partial h_\mathrm{sb}}\Biggr|_{h_\mathrm{sb}=0} = \int_0^{1/T}d\tau\,\bigl\langle m^\dag(\tau)m(0)\bigr\rangle,
    \label{eq:susc_qmc}
\end{equation}
with $m(\tau)=e^{-\tau\mathcal{H}}me^{\tau\mathcal{H}}$ in the Heisenberg picture where $\tau$ is the imaginary time. The right-hand side of Eq.~\eqref{eq:susc_qmc} is derived from the Kubo formula~\cite{PhysRevB.43.5950}. In the ferromagnetic XY model, $m=\sum_j(S_j^x+iS_j^y)/N$, while in the antiferromagnetic Heisenberg model, one has $m=\sum_j(-1)^jS_j^z/N$.

\subsubsection{The spin-$1/2$ ferromagnetic XY chain}

\begin{figure}[t!]
    \includegraphics[angle=0,clip=true,width=\columnwidth]{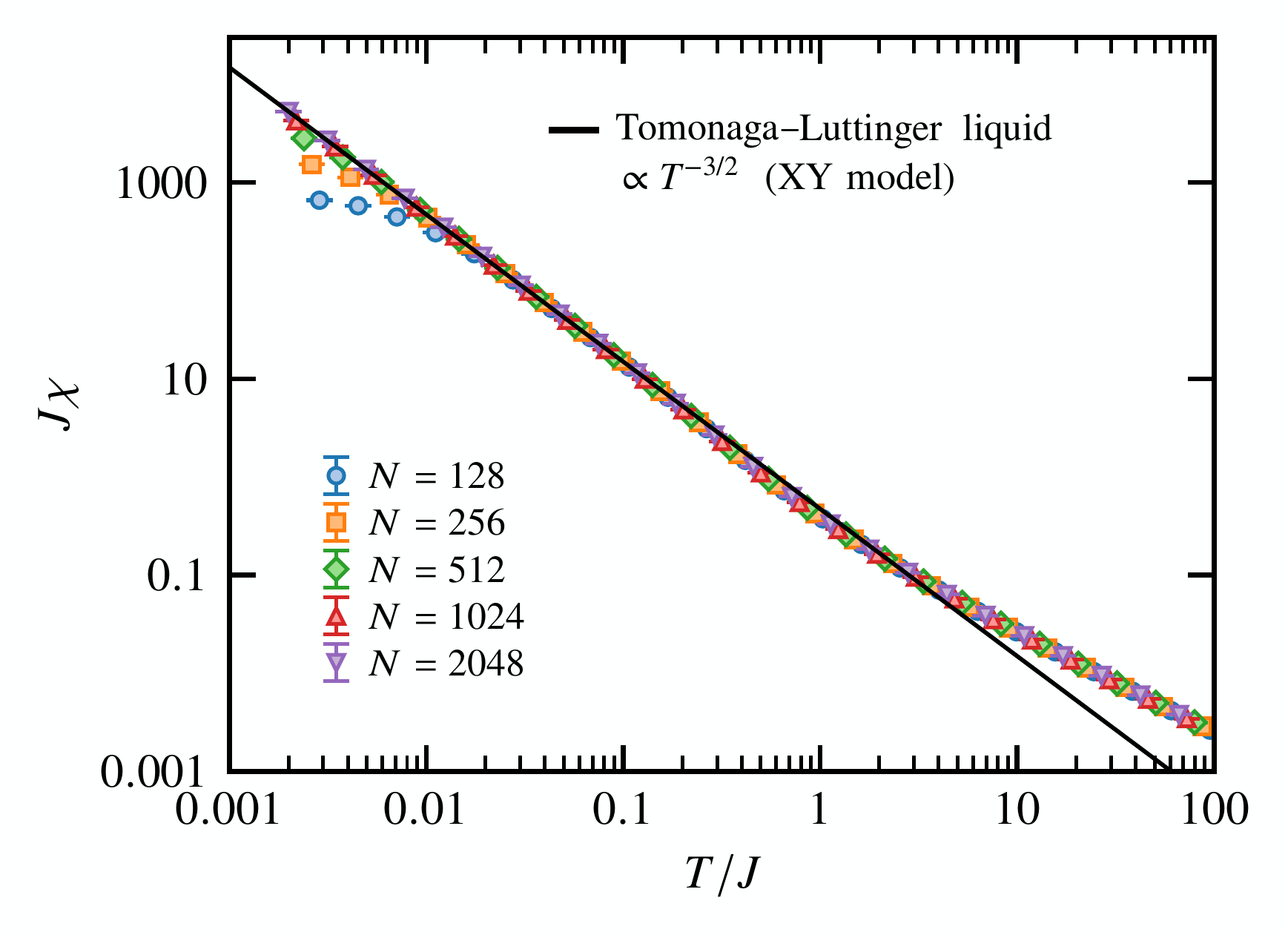}
    \caption{QMC transverse susceptibility of the XY chain ($\Delta=0$ and $p=0$) as a function of the temperature $T$ for different system sizes from $N=128$ to $N=2048$. The straight black line is the parameter-free expression of Eq.~\eqref{eq:1dchi_tll}. It fits asymptotically well the QMC data at low-energy for $T/J\lesssim 0.1$, when the Tomonaga-Luttinger liquid description becomes valid.}
    \label{fig:susc_xy}
\end{figure}

Equqation~\eqref{eq:1dchi_tll} provides a parameter-free expression, which gives for $\Delta=0$ in the low-temperature limit,
\begin{equation}
    J\chi_{xx}^{\Delta=0}(T)=0.474061...\left(\frac{J}{T}\right)^{3/2}.
    \label{eq:XY}
\end{equation}
This expression is plotted in Fig.~\ref{fig:susc_xy}, together with QMC results where one sees a very good agreement at low temperature.

\subsubsection{The spin-$1/2$ antiferromagnetic Heisenberg chain}
\label{sec:s12_af_chi}

\begin{figure}[t!]
    \includegraphics[angle=0,clip=true,width=\columnwidth]{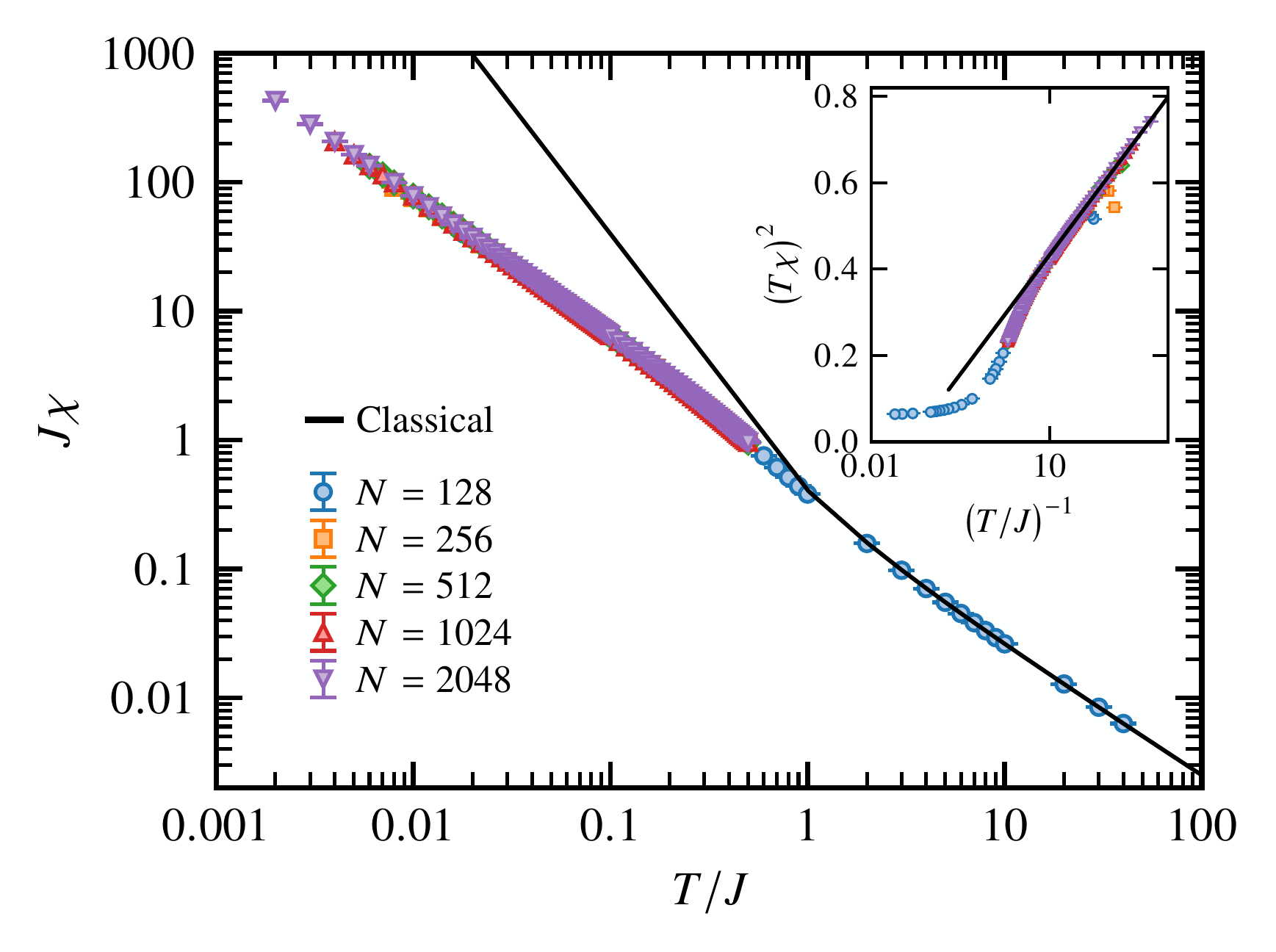}
    \caption{QMC results for the staggered susceptibility of the Heisenberg chain ($p=0$) as a function of the temperature $T$. Different symbols show system sizes from $N=128$ to $N=2048$. The classical result~\cite{fisher1964} is also shown by the solid line. Inset: $\bigl(T\chi\bigr)^2$ is plotted against $\bigl(T/J\bigr)^{-1}$. The solid line is a fit to the form $\chi_0^2\left(\ln \Lambda-\ln T\right)$, according to Eq.~\eqref{eq:susc_xxx}, with $\chi_0=0.2823(16)$ and $\Lambda=22.7(20)$.}
    \label{fig:susc_xxx}
\end{figure}

The Heisenberg spin-$1/2$ chain model is known to have logarithmic corrections in most observables~\cite{Affleck_1989,PhysRevB.43.8217,PhysRevLett.73.332,PhysRevB.56.13681,Affleck_1998,PhysRevB.63.140412,Barzykin_2000,PhysRevB.94.144409}. In particular, the staggered susceptibility~\footnote{In the SU(2) symmetric case, all spin orientations are equivalent.} is expected to follow~\footnote{Note that higher order corrections have also been computed analytically~\cite{PhysRevB.63.140412}, $\chi^{\Delta=1}_{\pi}(T)=\frac{\chi'_0}{T}\sqrt{\ln(J\Lambda'/T)+\ln\sqrt{\ln(J\Lambda'/T)}}$, but our data are best described by the simpler form Eq.~\eqref{eq:susc_xxx}},
\begin{equation}
    \chi^{\Delta=1}_{\pi}\bigl(T\bigr)=\frac{\chi_0}{T}\sqrt{\ln\bigl(J\Lambda/T\bigr)}.
    \label{eq:susc_xxx}
\end{equation}
In order to apply the RPA analysis, it appears very important to have a correct description for $\chi^{\Delta=1}_{\pi}(T)$. In Fig.~\ref{fig:susc_xxx} we show our QMC results for large spin chains, up to $N=2048$ sites. Our data are very well described by Eq.~\eqref{eq:susc_xxx} with $\chi_0=0.2823(16)$ and $\Lambda=22.7(20)$, in the temperature range $0.002J\le T\le 0.1J$. These parameters differ from the ones reported in Refs.~\cite{PhysRevB.55.14953,Kim:1998aa,PhysRevB.58.9142} where QMC was performed at higher temperature.

\section{Quantum Monte Carlo results for the small-world}
\label{sec:qmc}

We now turn to SW networks. In the undiluted case shown in Fig.~\ref{fig:lattices}(b), we average the QMC results over different lattices with $p>0$ (typically a few hundreds), since the long-ranged links are randomly drawn, while only one sample is enough for the disorder-free Hastings model of Fig.~\ref{fig:lattices}(c).

\subsection{Observables}

To characterize the finite-temperature transition, we consider the square of the order parameter, $\bigl\langle m^2\bigr\rangle$, directly accessible from the normalized structure factor for both the staggered antiferromagnetic Heisenberg and ferromagnetic XY models (see also Sec.~\ref{sec:susc_qmc}). It can also be evaluated by looking at the spin-spin correlation at long distance:
\begin{equation}
    \bigl\langle m^2\bigr\rangle = \lim_{|m-n|\to+\infty}\left\{
    \begin{array}{lr}
        \Bigl|\bigl\langle S^{z}_mS^{z}_n\bigr\rangle\Bigr|,\quad&\mathrm{XXX~case}\\
        \bigl\langle S^{x}_mS^{x}_n + S^{y}_mS^{y}_n\bigr\rangle,\quad&\mathrm{XY~case}.
    \end{array}
    \right.
    \label{eq:corr_longdist}
\end{equation}
On a finite-size system, the longest distance is taken along the $1$D ring with $|m-n|=N/2$. One can average over the $N/2$ pairs of such lattice sites.

Another quantity of interest is the fourth-order Binder ratio~\cite{PhysRevLett.47.693},
\begin{equation}
    Q = \bigl\langle m^4\bigr\rangle\Bigl/\bigl\langle m^2\bigr\rangle^2,
    \label{eq:binder_ratio}
\end{equation}
which takes a system-size independent value at the transition and is therefore useful to detect it.

\subsection{Mean-field behavior}

\subsubsection{Critical exponents}

In infinitely coordinated systems, where each site is coupled to all others (e.g., the Hastings model), the concepts of dimensionality and length, involved in the standard finite-size scaling hypothesis, are not well defined. Botet, Jullien, and Pfeuty extended the hypothesis to such systems~\cite{PhysRevLett.49.478,PhysRevB.28.3955} by substituting the correlation length $\xi$ with a coherence number $\mathcal{N}$, independent of the dimensionality. Similarly to $\xi$, it diverges at the transition $\mathcal{N}\sim|T-T_\mathrm{c}|^{-\tilde{\nu}}$ with $\tilde{\nu}$ a critical exponent depending on the system but not its dimension. The authors found that
\begin{equation}
    \tilde{\nu}=\nu_\mathrm{MF}\,d_\mathrm{u},
    \label{eq:nu_exponent}
\end{equation}
with $d_\mathrm{u}$ the upper critical dimension and $\nu_\mathrm{MF}$ the correlation length exponent of the MF theory. Equation~\eqref{eq:nu_exponent} has been verified for various physical systems and found to apply to more generic infinite-dimensional geometries such as SW networks~\cite{PhysRevLett.49.478,PhysRevB.28.3955,kim_xy_2001,yi_effect_2003}. For the XY and Heisenberg universality classes considered in this work, one has $d_\mathrm{u}=4$ and $\nu_\mathrm{MF}=1/2$, yielding $\tilde{\nu}=2$. As a result, close to the critical temperature $T_\mathrm{c}$, the square of the order parameter follows:
\begin{equation}
    \bigl\langle m^2\bigr\rangle = N^{-{2\beta_\mathrm{MF}}/{\tilde{\nu}}}\,\mathcal{F}_{m^2}\left(tN^{{1}/{\tilde{\nu}}}\right),
    \label{eq:m2_scaling}
\end{equation}
with $N$ the number of lattice sites, $\beta_\mathrm{MF}=1/2$ the order parameter MF exponent, $\mathcal{F}_{m^2}$ a universal scaling function, and $t=(T-T_\mathrm{c})/T_\mathrm{c}$ the reduced temperature. The Binder ratio of Eq.~\eqref{eq:binder_ratio} equally follows:
\begin{equation}
    Q = \mathcal{F}_{Q}\left(tN^{{1}/{\tilde{\nu}}}\right),
    \label{eq:Q_scaling}
\end{equation}
with $\mathcal{F}_{Q}$ the corresponding scaling function.

\subsubsection{Corrections to scaling}
\label{sec:corr_scaling}

\begin{figure}[t!]
    \includegraphics[width=1.0\columnwidth]{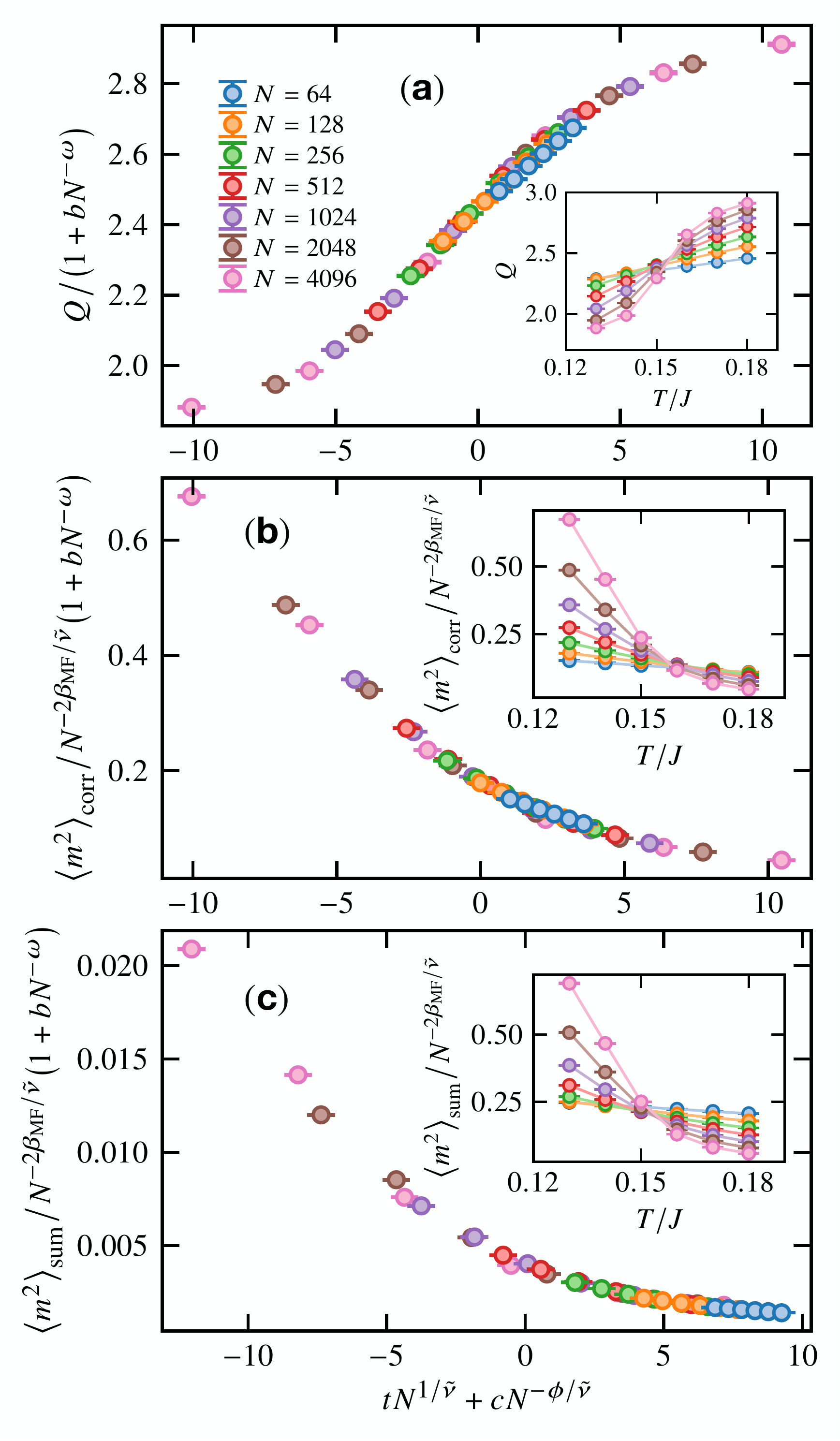}
    \caption{Hastings model. Data collapse for the staggered XXX antiferromagnet with $J'/J=0.25$. Three estimates are considered: (a) the binder cumulant, (b) the square of the order parameter evaluated from the spin-spin correlation at long distance, and (c) the normalized structure factor. Setting $\tilde{\nu}=2$ and $\beta_\mathrm{MF}=1/2$, we find that $T_\mathrm{c}\simeq 0.16$. The other fitting parameters are reported in Appendix~\ref{app:fits}. The insets show the data crossing without any correction to the scaling.}
    \label{fig:universality_hastings}
\end{figure}

Note that irrelevant corrections to the above scaling laws should also be considered with a modified scaling function,
\begin{equation}
    \mathcal{F}\to\bigl(1+bN^{-\omega}\bigr)\mathcal{F}\left(tN^{1/\tilde{\nu}}+cN^{-\phi/\tilde{\nu}}\right),
    \label{eq:scaling_correction}
\end{equation}
where $b$, $c$ are nonuniversal parameters, and $\omega,\,\phi$ are corrections to scaling exponents~\cite{Beach2005,wang2006}. In practice, we make a fourth-order Taylor expansion of the scaling function, i.e., $\mathcal{F}(x)\simeq\sum_{n=0}^4 a_nx^n$ with $x=tN^{1/\tilde{\nu}}+cN^{-\phi/\tilde{\nu}}$ according to Eq.~\eqref{eq:scaling_correction}. At this stage, $T_\mathrm{c},\,b,\,c,\,\omega,\,\phi,\,a_0,\,a_1,\,a_2,\,a_3$, and $a_4$ are all parameters obtained by nonlinear least squares fitting. Using the values of the exponents $\beta_\mathrm{MF}$ and $\tilde{\nu}$ give very good data collapses (see Fig.~\ref{fig:universality_hastings}).

We use a standard least-squares fitting method to obtain the parameters. For each dataset, the fitting procedure is repeated $\approx 10^3$ times where each data point is generated from a normal distribution of mean and standard deviation corresponding to the statistical QMC average and error, respectively~\footnote{For the SW network, the statistical QMC average result of a given simulation is averaged over randomly drawn lattice geometries. Note also that because the fitting form is relatively complicated, we have found that it can be advantageous to perform, at first, the fit on the scaling forms without the corrections to the scaling. We then use the obtained fitting parameters as initial guesses for the more complicated form. In all cases, we add a small random noise to each initial guess for the parameters.}.

\subsubsection{Hastings model}

We first discuss the infinitely connected Hastings model where the long-range couplings take the form $|J_{ij}^\mathrm{LR}|=J'/N$, $\forall |i-j|>1$. Figure~\ref{fig:universality_hastings} shows QMC results for the staggered antiferromagnetic Heisenberg model with $J'/J=0.25$. The three panels display the different observables used to extract the ordering transition, which all agree perfectly with an estimate $T_c\simeq 0.16$.

\begin{figure}
    \includegraphics[angle=0,clip=true,width=.975\columnwidth]{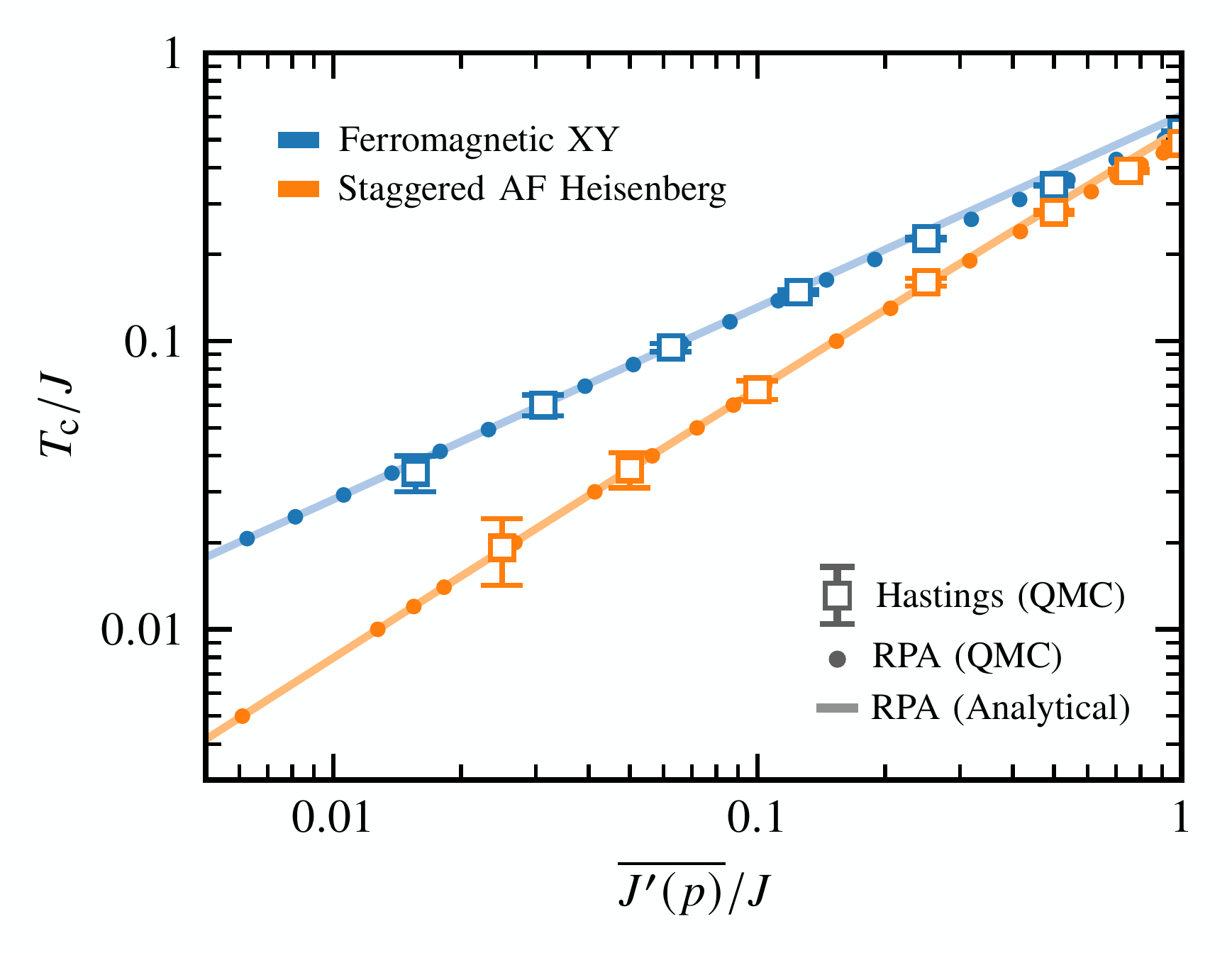}
    \caption{Critical temperature of the Hastings model, plotted as a function of the average long-range coupling strength $\overline{J'(p)}/J$ for the ferromagnetic XY and staggered antiferromagnetic Heisenberg models. The RPA estimates (analytical and QMC) are also displayed for each model.}
    \label{fig:tc_mf}
\end{figure}

The same analysis can be performed for various values of $J'/J$, for both the XY ferromagnet and the staggered XXX antiferromagnet. QMC results for the critical temperature as a function of $J'/J$ are reported in Fig.~\ref{fig:tc_mf} where a direct comparison to the RPA prediction is provided. Here, we clearly observe that not only the critical exponents obey MF predictions (Fig.~\ref{fig:universality_hastings}), but the chain-MF theory provides through the RPA prediction of Eq.~\eqref{eq:Tcrpa} a perfectly quantitative estimate for $T_\mathrm{c}$ of the Hastings model, and this remains true up to $J'/J=1$. Therefore the 1D character of the underlying spin-chain lattice is fundamental, despite the infinitely connected nature of the Hastings model.

$T_\mathrm{c}^\mathrm{RPA}$ is also shown in Fig.~\ref{fig:tc_mf}, together with QMC data. For the XY ferromagnet, Eq.~\eqref{eq:1dTC_tll} yields
\begin{equation}
    T_\mathrm{c,\,XY}^\mathrm{RPA}\bigr/J= 0.60798...\left(\frac{J'}{J}\right)^{2/3},
\end{equation}
which clearly agrees very well with QMC below $J'/J\sim 0.2$. The staggered XXX antiferromagnetic case is a bit more subtle because of logarithmic corrections discussed above. Interestingly, the rather simple form of Eq.~\eqref{eq:susc_xxx} yields the following RPA expression for the critical temperature:
\begin{equation}
    \frac{\chi_0}{T_\mathrm{c}}\sqrt{\ln(J\Lambda/T_\mathrm{c})}=\frac{1}{J'}
\end{equation}
Using the Lambert $W$ function~\cite{corless_lambertw_1996}, it gives, for  $J'\ll J$,
\begin{equation}
    T^{\mathrm{RPA}}_\mathrm{c,\,XXX}\approx\chi_0J'\sqrt{\ln\left(\frac{J}{J'}\right)+A-\ln\sqrt{2\ln\left(\frac{J}{J'}\right)+2A}},
    \label{eq:tc_rpa_analytical}
\end{equation}
with $\chi_0=0.2823(16)$ and $A=\ln\left(\Lambda\sqrt{2}\right)-\ln\chi_0=4.73(9)$. This RPA analytical estimate for $T_\mathrm{c}$ compares very well to QMC data, as shown in Fig.~\ref{fig:tc_mf}.

In this toy model with infinite connectivity, the chain-MF theory provides with the RPA expression an exact estimate for the critical temperature measured by QMC. As anticipated by Hastings~\cite{Hastings2003}, the simple MF expression [Eq.~\eqref{eq:TcMF}] obtained for SW geometries in the diluted limit does not apply in this case. In the following we address the disordered case with low connectivity for both the XY ferromagnet and the XXX antiferromagnet.

\subsection{The Undiluted small-world geometry}

\begin{figure}[!t]
    \includegraphics[width=0.99\columnwidth]{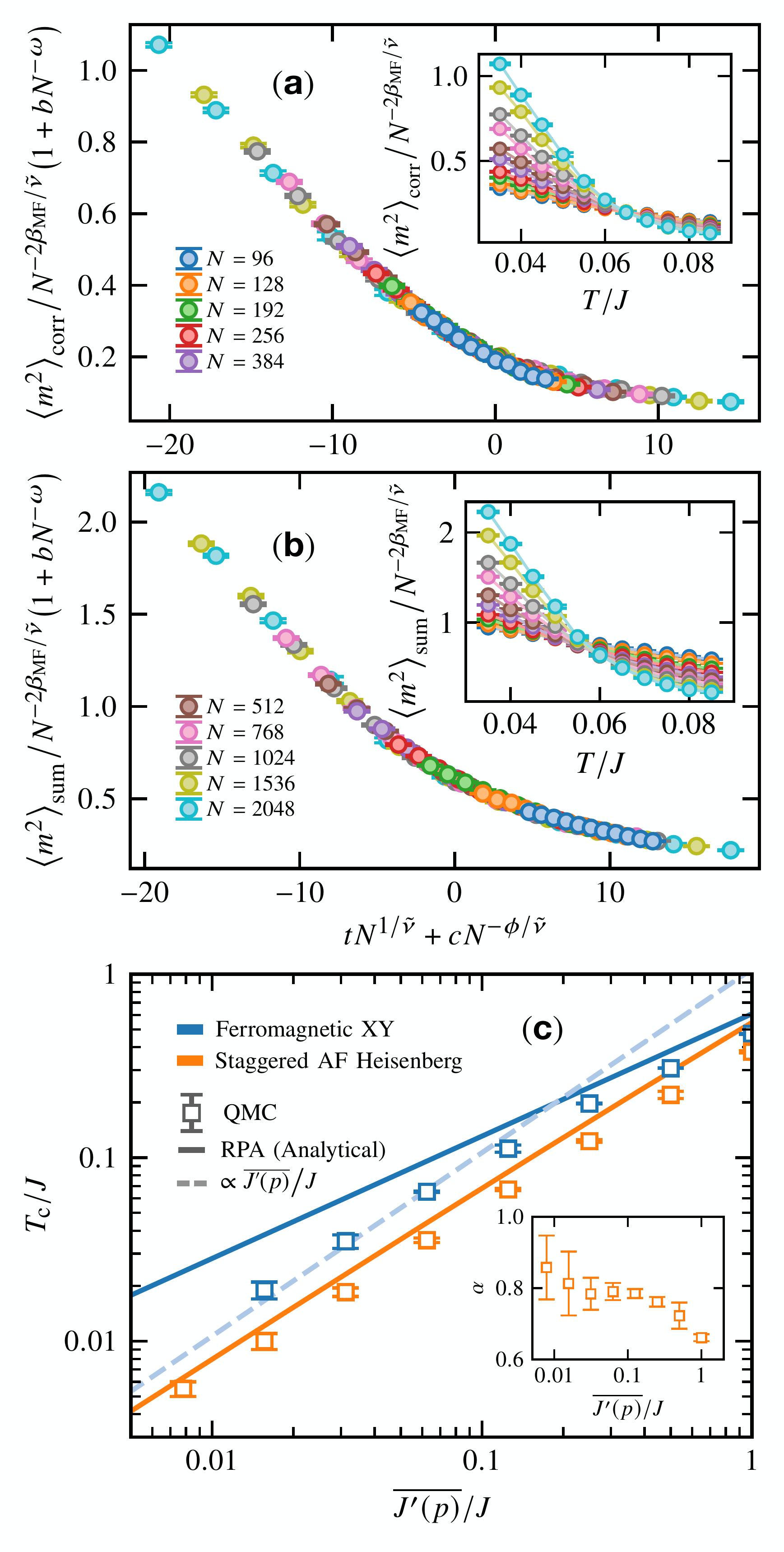}
    \caption{Data collapse for the XY ferromagnet on the SW network with $p=0.03125$: the square of the order parameter evaluated from (a) the normalized structure factor and (b) the spin-spin correlation at long distance. Setting $\tilde{\nu}=2$ and $\beta_\mathrm{MF}=1/2$, we find that $T_\mathrm{c}\simeq 0.065$. The other fitting parameters are reported in Appendix~\ref{app:fits}. The insets show the data crossing without any correction to the scaling. (c) Critical temperature of the ferromagnetic XY and staggered antiferromagnetic Heisenberg models versus the average strength of the extra couplings. In each case, the analytical RPA estimate is also displayed. In the staggered antiferromagnetic Heisenberg model, the two estimates agree as $\overline{J'(p)}\to 0$. We plot in the inset the renormalization parameter $\alpha$ (see text). As $\overline{J'(p)}\to 0$, one sees that $\alpha\to 1$. In the ferromagnetic XY case, the RPA and QMC estimates deviate below $\overline{J'(p)}/J\sim 0.1$. The dashed line is a linear fit $\propto\overline{J'(p)}/J$.}
    \label{fig:universality_sw}
\end{figure}

We now turn to the disordered case with a finite branching probability $p\le 0.5$ and long-range couplings $J'=J$. In contrast to the disorder-free Hastings model, here we have to perform disorder averaging, typically over a few hundreds of independent samples. The average distance between shortcuts being $\zeta_p\approx (2p)^{-1}$, QMC simulations have to be ideally achieved over systems of length $N\gg \zeta_p$. This natural scale fixes a limit to the accessible concentrations $p\gtrsim 10^{-2}$ in our simulations.

Despite the very low connectivity ${\overline{z}}=2+2p$, a finite temperature transition is clearly detected in our QMC simulations (see Fig.~\ref{fig:universality_sw}). We obtain MF critical exponents, as expected from the $d=\infty$ nature of the SW network, even in the vanishing $p$ limit. Nevertheless, there are notable differences with the infinitely connected Hastings model, as we discuss now.

Figure~\ref{fig:universality_sw}(c) shows the concentration $p$ dependence of the critical temperature for both ferromagnetic (XY) and antiferromagnetic (XXX) ordering transitions. QMC estimates for $T_\mathrm{c}$ are compared to the RPA result. We first discuss the staggered XXX antiferromagnet (orange). In this case, QMC results and RPA estimates are not equal, but they seemingly get closer when ${\overline{J'(p)}}\to 0$. Following similar ideas developed in Refs.~\cite{PhysRevB.61.6757,Yasuda2005,PhysRevB.74.184407,thielemann_field-controlled_2009,bollmark_dimensional_2020}, we introduce a renormalization parameter $\alpha$, such that the exact critical temperature follows from the RPA formula, with a $p$-dependent renormalization of the average long-range coupling $\overline{J'(p)}$,
\begin{equation}
    T_\mathrm{c}=\chi^{-1}\left(\frac{1}{\alpha {\overline{J'(p)}}}\right).
    \label{eq:Tcrpa_alpha}
\end{equation}
We observe in the inset of Fig.~\ref{fig:universality_sw}(c) that $\alpha$ seems to increase towards unity when $p\to 0$. While it is difficult to draw a definite conclusion, if confirmed, this would make the RPA result asymptotically exact in this extreme limit.

The XY model shows a strikingly different trend. Indeed, while the RPA behavior~$T_\mathrm{c}\sim ({\overline{J'(p)}}/J)^{2/3}$ gives a reasonable description of the exact QMC data at intermediate coupling strengths, this is no longer the case when ${\overline{J'(p)}}/J\lesssim 0.1$, where an increasing deviation is clearly observed. This result is a consequence of the crossover from RPA to MF discussed in Sec.~\ref{sec:crossover}. Indeed, taking Eq.~\eqref{eq:pstar} for the XY case at $\Delta=0$, we anticipate a crossover probability $p^\star\approx 0.11$ below which the average distance between two shortcuts $\zeta_p$ becomes larger than the 1D correlation length value at the RPA temperature $\xi(T_\mathrm{c}^\mathrm{RPA})$. As predicted, we clearly observe a downturn for $T_\mathrm{c}$ towards the linear behavior in Eq.~\eqref{eq:1DMF} shown by a dashed line in Fig.~\ref{fig:universality_sw}(c). This crossover is a direct consequence of the underlying Luttinger liquid behavior which allows to break the condition of Eq.~\eqref{eq:condition}: here $\eta>1$, with $\eta=\arccos(-\Delta)/\pi\le 1$ for the $d=1$ XXZ model.

Note that for the XXX case, the RPA estimate in Eq.~\eqref{eq:tc_rpa_analytical} has multiplicative logarithmic corrections, slowly growing when $p\to 0$. Therefore one should expect, in principle, to observe a similar crossover towards the MF expression [Eq.~\eqref{eq:1DMF}] for the staggered XXX antiferromagnet. However, this effect is clearly out of reach since it would theoretically occur for $p^\star\approx 10^{-9}$.

\subsection{Influence of a spin gap}

\begin{figure}[!t]
    \includegraphics[width=\columnwidth]{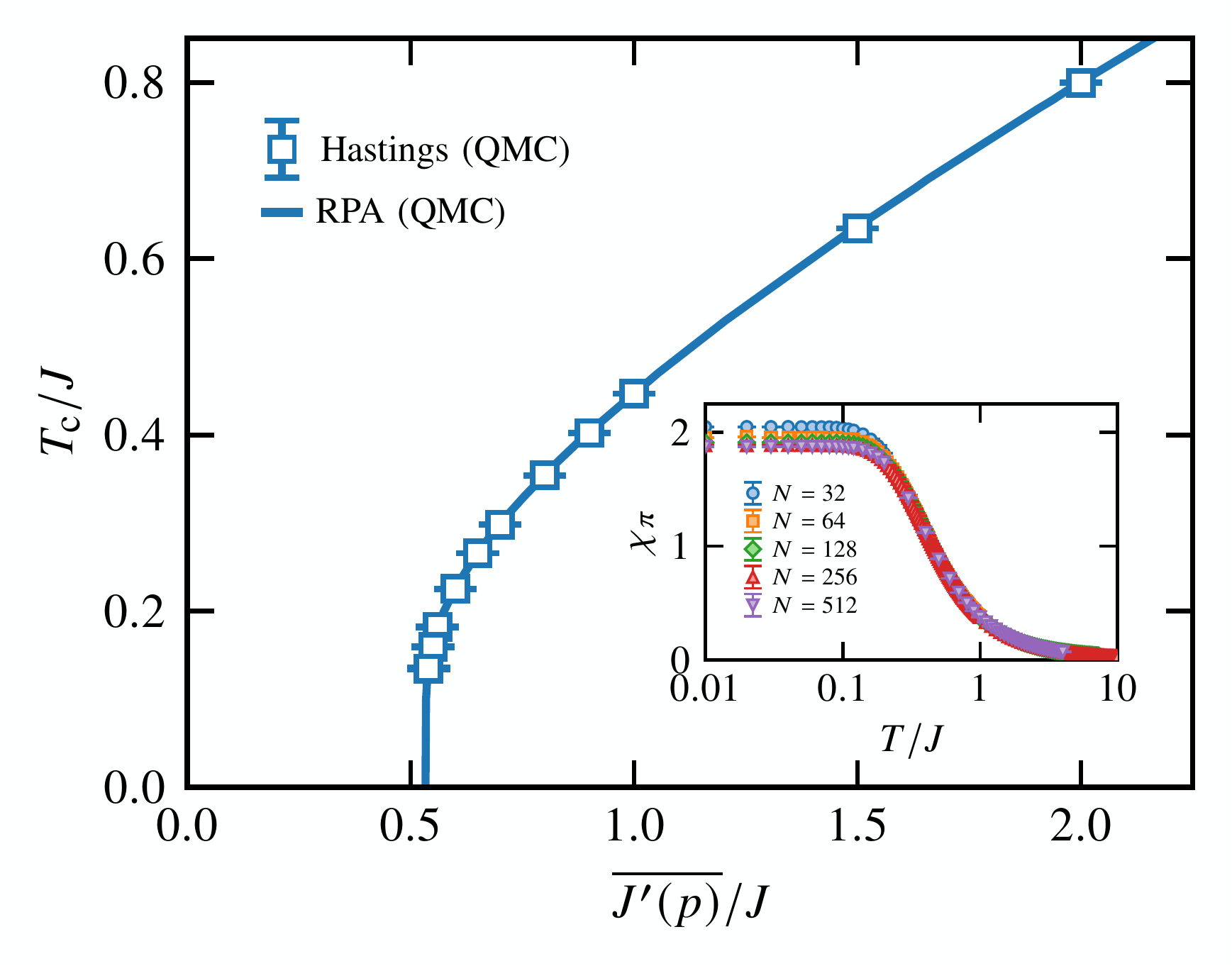}
    \caption{Critical temperature of the dimerized antiferromagnetic model, Eq.~\eqref{eq:dim}, with $\delta=0.25$ and additional long-range couplings of the Hastings form of strength $\overline{J'(p)}$. The RPA estimate compares perfectly to the QMC results. Inset: Temperature dependence of the staggered susceptibility of the $1$D dimerized system.}
    \label{fig:tc_gap}
\end{figure}

We finally investigate a dimerized antiferromagnetic chain, governed by the following Heisenberg Hamiltonian,
\begin{equation}
    \mathcal{H}_\mathrm{1D}=J\sum\nolimits_{i=1}^{N}\left[1+\delta(-1)^i\right]{\boldsymbol{S}}_{i}\cdot{\boldsymbol{S}}_{i+1}.
    \label{eq:dim}
\end{equation}
In contrast with the previous study, here the ring  has a gapped ground state, with a finite $T=0$ correlation length~\cite{PhysRevB.19.402}. Instead of a divergent staggered susceptibility, now $\chi_{\pi}$ saturates at low temperature to a finite value, $\chi_{\pi}^{0}$, as visible in Fig.~\ref{fig:tc_gap} (inset) for a dimerization parameter $\delta=0.25$.

Consequently, according to the RPA, Eq.~\eqref{eq:Tcrpa}, the absence of low-$T$ divergence for $\chi_\pi$ should imply a critical coupling $J'_\mathrm{c}=1/\chi_{\pi}^{0}$, below which $T_\mathrm{c}=0$. This is well known for instance in the case of coupled Haldane chains~\cite{PhysRevB.42.4537,PhysRevLett.112.247203}. Here we performe QMC simulations of the Hastings model with extra couplings of varying strength $J'/N$ for a dimerization parameter $\delta=0.25$. $T_\mathrm{c}$ estimates are reported in Fig.~\ref{fig:tc_gap}, together with the RPA result, obtained using $T_\mathrm{c}^\mathrm{RPA}=\chi_{\pi}^{-1}(1/J')$ when a solution exists, and $T_\mathrm{c}=0$ otherwise. The agreement is excellent, even when $J'/J>1$. The $T=0$ critical coupling $J'_\mathrm{c}\approx 0.53$ is also perfectly captured by the RPA treatment.

We expect qualitatively the same physics on the undiluted small-world network in the presence of a spin gap: there will be a finite critical temperature $T_\mathrm{c}$ above a critical value of the strength of the extra couplings (and zero below). However, we do not expect the QMC and RPA results to perfectly agree, as they do on the Hastings geometry (see Fig.~\ref{fig:tc_gap}). The RPA treatment will overestimate the exact QMC results, as in the small-world network built against 1D gapless spin chains. Based on this, we also expect the agreement between RPA and QMC to become better for smaller values of the spin gap, which will lead to a finite $T_\mathrm{c}$ in the limit $\overline{J'(p)}/J\to 0$.

\section{Summary and conclusion}
\label{sec:conclusion}

In this work, building on mean-field theory and extensive quantum Monte Carlo simulations, we investigated interacting quantum spins on small-world networks. Starting from $1$D rings, we considered two situations: all-to-all interacting and long-range interactions randomly added. The effective infinite dimension of the lattice leads to a magnetic ordering at finite temperature $T_\mathrm{c}$ with mean-field criticality.

First, we showed that different mean-field treatments led to different power-law behaviors for the scaling of $T_\mathrm{c}$ versus the average strength $\overline{J'}$ of the extra couplings. The first approach is controlled by a competition between a characteristic length scale of the small-world network and the thermal correlation length of the underlying 1D system, and leads to $T_\mathrm{c}\propto\overline{J'}$. The other approach is based on the random phase approximation. For a critical $1$D system with anomalous exponent $\eta$, it gives $T_\mathrm{c}\propto\overline{J'}^{1/(2-\eta)}$.

Before confronting these approximate treatments with unbiased quantum Monte Carlo simulations of the problem, we compared analytical RPA based on low-energy physics with numerical RPA in order to quantitatively define the low-temperature limit of the analytical approaches. By computing the transverse susceptibility of the XY chain and the staggered susceptibility of the Heisenberg chain with exchange coupling $J$, we found that the analytical low-energy approaches become asymptotically exact for $T\lesssim J$.

Starting with the all-to-all interacting system (Hastings model), we first checked that the transition belonged to the mean-field universality class. Because of the effective infinite dimensionality of the system, the correlation length exponent is rescaled as $\tilde{\nu}\to\nu_\mathrm{MF}d_\mathrm{u}$ with $\nu_\mathrm{MF}=1/2$ the standard mean-field theory exponent and $d_\mathrm{u}=4$ the upper critical dimension~\cite{PhysRevLett.49.478,PhysRevB.28.3955}, which we verified in a finite-size scaling analysis. Finally, we found that for the Hastings model, the critical temperature $T_\mathrm{c}$ scales according to the random phase approximation versus the average strength $\overline{J'}$ of the extra couplings.

We then considered the system with long-range interactions, randomly added with a finite probability $p$. Similarly to the Hastings model, we checked that its criticality belongs to the mean-field universality class. However, we found in this case that the critical temperature shows both the $T_\mathrm{c}\propto\overline{J'}^{1/(2-\eta)}$ and $T_\mathrm{c}\propto\overline{J'}$ scalings, with a crossover from one to the other. As $\overline{J'}$ is reduced, the linear scaling takes over the RPA behavior when the characteristic length scale of the network becomes larger than the 1D thermal correlation length at the RPA temperature.

Finally, we investigated the fate of a gapped 1D spin chain against the small-world effect by considering the dimerized spin-half Heisenberg chain. We found that the gap of the $1$D system leads to a critical value $\overline{J'_\mathrm{c}}$ for magnetic ordering. Beyond $\overline{J'_\mathrm{c}}$, the critical temperature behavior is well captured by the RPA estimate.

For future work, it would be interesting to investigate how the order parameter at zero temperature responds to the small-world effect. For instance, a TLL-based approach predicts that $\langle m^2\rangle\propto \overline{J'}^{\eta/(2-\eta)}$~\cite{giamarchi2003quantum,PhysRevLett.101.137207,bouillot_statics_2011,PhysRevB.94.144403,blinder_nuclear_2017}, but as for the scaling of the critical temperature, one might expect a crossover towards another MF regime as a function of $\overline{J'}$.

Besides, it is very stimulating to think of the small-world effect in the presence of disorder~\cite{PhysRevE.72.036203}. It has been found to host unusual physics for noninteracting fermions~\cite{PhysRevLett.118.166801,PhysRevResearch.2.012020}, and it would be interesting to study the problem for interacting quantum systems, similarly to what has been done on the Cayley tree~\cite{PhysRevB.102.174205} for bosons in a random potential. Random exchange spin systems also offer a very promising platform, in particular to explore the issue of infinite randomness criticality~\cite{PhysRevB.51.6411} against the small-world effect~\cite{PhysRevE.72.066101}. We further note that magnetic frustration, occurring in the long-range interactions across the ring~\cite{PhysRevLett.104.137204} for instance, is another fascinating route where one could find more exotic quantum phases of matter. Finally, it is fascinating to observe that small-world quantum magnets are now available in experiments, with for instance all-to-all spin models, or tree like tunable Heisenberg-type systems which can be realized in cold-atom setups coupled to an optical cavity~\cite{PhysRevX.9.041011,PhysRevLett.123.130601,PhysRevLett.125.060402}.

\begin{acknowledgments}
    We are grateful to G. Lemari\'e for very fruitful discussions. M.D. was supported by the U.S. Department of Energy, Office of Science, Office of Basic Energy Sciences, Materials Sciences and Engineering Division under Contract No. DE-AC02-05-CH11231 through the Scientific Discovery through Advanced Computing (SciDAC) program (KC23DAC Topological and Correlated Matter via Tensor Networks and Quantum Monte Carlo). N. L. acknowledges the French National Research Agency (ANR) under Projects THERMOLOC ANR- 16-CE30-0023-02, and GLADYS ANR-19-CE30-0013. This research used the Lawrencium computational cluster resource provided by the IT Division at the Lawrence Berkeley National Laboratory (supported by the Director, Office of Science, Office of Basic Energy Sciences, of the U.S. Department of Energy under Contract No. DE-AC02-05CH11231). This research also used resources of the National Energy Research Scientific Computing Center (NERSC), a U.S. Department of Energy Office of Science User Facility operated under Contract No. DE-AC02-05CH11231. We also acknowledge the use of HPC resources from CALMIP (Grants No. 2019-P0677 and No. 2020-P0677) and GENCI (Grant No. x2020050225).
\end{acknowledgments}

\appendix
\section{Chain mean-field theory and random phase approximation}
\label{app:RPA}

\subsection{Chain mean-field theory}

We recall the basic idea of treating the long-range part in mean field. One looks at the fluctuations around the average value of the operators in the long-range part by making the substitution
\begin{equation}
    S^{x,y,z}_i=\bigl\langle S^{x,y,z}_i\bigr\rangle + \Bigl(S^{x,y,z}_i - \bigl\langle S^{x,y,z}_i\bigr\rangle\Bigr).
\end{equation}
Neglecting quadratic terms and up to an irrelevant constant, one gets the following effective 1D Hamiltonian,
\begin{equation}
    {\cal{H}}_\mathrm{eff}^\mathrm{XXX}=J\sum\nolimits_i {\boldsymbol{S}}_i\cdot {\boldsymbol{S}}_{i+1}+\sum\nolimits_{i,j} J^\mathrm{LR}_{ij}\bigl\langle S_j^z\bigr\rangle S_i^z,
\end{equation}
in the staggered antiferromagnetic Heisenberg case ($\Delta=1$), where magnetic order, induced by long-range couplings, has been assumed along the $z$ spin component. For the ferromagnetic XY model ($\Delta=0$), the ordering is expected in the $xy$ plane. We suppose it is along the $x$ spin component and obtain the effective 1D model as follows:
\begin{equation}
    {\cal{H}}_\mathrm{eff}^\mathrm{XY}=J\sum\nolimits_i \left(S_i^x S_{i+1}^{x}+S_i^y S_{i+1}^{y}\right)+\sum\nolimits_{i,j}J^\mathrm{LR}_{ij}\bigl\langle S_j^x\bigr\rangle S_i^x.
\end{equation}
In both cases, the idea is to assume symmetry breaking. Considering a homogeneous system, one can replace $\bigl\langle S_j^{x,z}\bigr\rangle$ by the corresponding order parameter $\langle m\rangle$. In the staggered antiferromagnetic Heisenberg case,
\begin{equation}
    {\cal{H}}_\mathrm{eff}^\mathrm{XXX}=J\sum\nolimits_i {\boldsymbol{S}}_i\cdot {\boldsymbol{S}}_{i+1}+\langle m\rangle\sum\nolimits_{i,j}J_{ij}^\mathrm{LR} (-1)^{j}S_i^z,
    \label{eq:Heff1D_xxx}
\end{equation}
where the factor $(-1)^j$ comes from the fact that $\langle m\rangle=(-1)^j\bigl\langle S_j^z\bigr\rangle$. In the XY case, the order is ferromagnetic:
\begin{equation}
    {\cal{H}}_\mathrm{eff}^\mathrm{XY}=J\sum\nolimits_i \left(S_i^x S_{i+1}^{x}+S_i^y S_{i+1}^{y}\right)+\langle m\rangle\sum\nolimits_{i,j}J_{ij}^\mathrm{LR}S_i^x.
    \label{eq:Heff1D_xy}
\end{equation}

\subsection{Random phase approxmation}

The linear response to a tiny symmetry-breaking field $h_\mathrm{sb}$ coupled to the order parameter operator $m$ takes the form
\begin{equation}
    \langle m\rangle=\chi^\mathrm{1D}\bigl(T\bigr)\,h_\mathrm{sb},
\end{equation}
where $\chi^\mathrm{1D}(T)$ is the susceptibility. Neglecting possible inhomogeneities in the SW branching by using the fact that the average strength of extra couplings across the ring is $\overline{J'(p)}=2pJ'$ and including explicitly the symmetry-breaking field for the antiferromagnetic XXX model of Eq.~\eqref{eq:Heff1D_xxx}, one gets,
\begin{eqnarray}
    {\cal{H}}_\mathrm{eff}^\mathrm{XXX}=&&J\sum\nolimits_i {\boldsymbol{S}}_i\cdot {\boldsymbol{S}}_{i+1}+\langle m\rangle\overline{J'(p)}\sum\nolimits_{i}(-1)^iS_i^z\nonumber\\
    &&+ h_\mathrm{sb}\sum\nolimits_i (-1)^{i}S_i^z,
\end{eqnarray}
and similarly for the XY case of Eq.~\eqref{eq:Heff1D_xy},
\begin{eqnarray}
    {\cal{H}}_\mathrm{eff}^\mathrm{XY}=&&J\sum\nolimits_i {\boldsymbol{S}}_i\cdot {\boldsymbol{S}}_{i+1}+\langle m\rangle\overline{J'(p)}\sum\nolimits_{i}S_i^x\nonumber\\
    &&+h_\mathrm{sb}\sum\nolimits_i S_i^x.
\end{eqnarray}
Within the chain mean-field approach, the total effective symmetry breaking field is $h_\mathrm{sb}+\overline{J'(p)}\langle m\rangle$. Therefore,
\begin{equation}
    \langle m\rangle=\chi^\mathrm{1D}(T)\Bigl(h_\mathrm{sb} + \overline{J'(p)}\langle m\rangle\Bigr).
    \label{eq:m1D}
\end{equation}
Isolating the order parameter from Eq.~\eqref{eq:m1D}, one gets,
\begin{equation}
    \langle m\rangle=\frac{\chi^\mathrm{1D}(T)}{1-\overline{J'(p)}\chi^\mathrm{1D}(T)}h_\mathrm{sb}=\chi^\mathrm{RPA}(T)h_\mathrm{sb}.
\end{equation}
Because magnetic ordering occurs at $T=T_\mathrm{c}^\mathrm{RPA}$ with a divergence of the susceptibility, one finds the condition,
\begin{equation}
    \chi^\mathrm{1D}\Bigl(T_\mathrm{c}^\mathrm{RPA}\Bigr)=1\Bigl/\overline{J'(p)}.
\end{equation}
By inverting it, one can get the RPA estimate of the critical temperature $T_\mathrm{c}^\mathrm{RPA}$.

\section{Fitting parameters for the scaling functions}
\label{app:fits}

Following the scaling analysis including corrections to the scaling (see Sec.~\ref{sec:corr_scaling}), the fitting parameters for the data collapse of Figs.~\ref{fig:universality_hastings} and ~\ref{fig:universality_sw} are reported in Table~\ref{tab:fits}.

\begin{table}[!h]
    \vspace*{5mm}
    \begin{minipage}{1.0\columnwidth}
        \center
        \begin{ruledtabular}
            \begin{tabular}{lccccc}
                \thead{Quantity} & \thead{$T_\mathrm{c}$} & \thead{$\omega$} & \thead{$\phi$} & \thead{$b$} & \thead{$c$}\\
                \hline
                \multicolumn{6}{c}{Staggered antiferromagnetic Heisenberg (Hastings, $J'/J=0.125$)}\\
                \hline
                \makecell{$Q$ (Binder)} & \makecell{$0.154(5)$} & \makecell{$1.49(2)$} & \makecell{$3.72(1)$} & \makecell{$-41(1)$} & \makecell{$4613(10)$}\\
                \makecell{$\langle m^2\rangle_\mathrm{corr}$} & \makecell{$0.156(4)$} & \makecell{$4.89(7)$} & \makecell{$0.66(2)$} & \makecell{$-1.60(3)$} & \makecell{$9.24(5)$}\\
                \makecell{$\langle m^2\rangle_\mathrm{sum}$} & \makecell{$0.165(7)$} & \makecell{$0.37(6)$} & \makecell{$0.7(1)$} & \makecell{$667(9)$} & \makecell{$37(1)$}\\
                \hline
                \multicolumn{6}{c}{XY Ferromagnet (small world, $p=0.03125$)}\\
                \hline
                \makecell{$\langle m^2\rangle_\mathrm{corr}$} & \makecell{$0.064(3)$} & \makecell{$1.67(3)$} & \makecell{$2.74(8)$} & \makecell{$81(2)$} & \makecell{$-30(1)$}\\
                \makecell{$\langle m^2\rangle_\mathrm{sum}$} & \makecell{$0.061(6)$} & \makecell{$1.20(8)$} & \makecell{$2.08(12)$} & \makecell{$290(4)$} & \makecell{$1036(12)$}
            \end{tabular}
        \end{ruledtabular}
    \end{minipage}
    \caption{Fitting parameters for the data collapse of Figs.~\ref{fig:universality_hastings} and ~\ref{fig:universality_sw}. See Sec.~\ref{sec:corr_scaling} for a definition of the different parameters.}
    \label{tab:fits}
\end{table}

\bibliography{sw}

\begin{thebibliography}{99}%
\makeatletter
\providecommand \@ifxundefined [1]{%
 \@ifx{#1\undefined}
}%
\providecommand \@ifnum [1]{%
 \ifnum #1\expandafter \@firstoftwo
 \else \expandafter \@secondoftwo
 \fi
}%
\providecommand \@ifx [1]{%
 \ifx #1\expandafter \@firstoftwo
 \else \expandafter \@secondoftwo
 \fi
}%
\providecommand \natexlab [1]{#1}%
\providecommand \enquote  [1]{``#1''}%
\providecommand \bibnamefont  [1]{#1}%
\providecommand \bibfnamefont [1]{#1}%
\providecommand \citenamefont [1]{#1}%
\providecommand \href@noop [0]{\@secondoftwo}%
\providecommand \href [0]{\begingroup \@sanitize@url \@href}%
\providecommand \@href[1]{\@@startlink{#1}\@@href}%
\providecommand \@@href[1]{\endgroup#1\@@endlink}%
\providecommand \@sanitize@url [0]{\catcode `\\12\catcode `\$12\catcode
  `\&12\catcode `\#12\catcode `\^12\catcode `\_12\catcode `\%12\relax}%
\providecommand \@@startlink[1]{}%
\providecommand \@@endlink[0]{}%
\providecommand \url  [0]{\begingroup\@sanitize@url \@url }%
\providecommand \@url [1]{\endgroup\@href {#1}{\urlprefix }}%
\providecommand \urlprefix  [0]{URL }%
\providecommand \Eprint [0]{\href }%
\providecommand \doibase [0]{http://dx.doi.org/}%
\providecommand \selectlanguage [0]{\@gobble}%
\providecommand \bibinfo  [0]{\@secondoftwo}%
\providecommand \bibfield  [0]{\@secondoftwo}%
\providecommand \translation [1]{[#1]}%
\providecommand \BibitemOpen [0]{}%
\providecommand \bibitemStop [0]{}%
\providecommand \bibitemNoStop [0]{.\EOS\space}%
\providecommand \EOS [0]{\spacefactor3000\relax}%
\providecommand \BibitemShut  [1]{\csname bibitem#1\endcsname}%
\let\auto@bib@innerbib\@empty
%</preamble>
\bibitem [{\citenamefont {Scott}(1988)}]{scott_social_1988}%
  \BibitemOpen
  \bibfield  {author} {\bibinfo {author} {\bibfnamefont {J.}~\bibnamefont
  {Scott}},\ }\href {\doibase 10.1177/0038038588022001007} {\bibfield
  {journal} {\bibinfo  {journal} {Sociology}\ }\textbf {\bibinfo {volume}
  {22}},\ \bibinfo {pages} {109} (\bibinfo {year} {1988})}\BibitemShut
  {NoStop}%
\bibitem [{\citenamefont {Watts}\ and\ \citenamefont
  {Strogatz}(1998)}]{watts_collective_1998}%
  \BibitemOpen
  \bibfield  {author} {\bibinfo {author} {\bibfnamefont {D.~J.}\ \bibnamefont
  {Watts}}\ and\ \bibinfo {author} {\bibfnamefont {S.~H.}\ \bibnamefont
  {Strogatz}},\ }\href {\doibase 10.1038/30918} {\bibfield  {journal} {\bibinfo
   {journal} {Nature}\ }\textbf {\bibinfo {volume} {393}},\ \bibinfo {pages}
  {440} (\bibinfo {year} {1998})}\BibitemShut {NoStop}%
\bibitem [{\citenamefont {Barab{\'a}si}\ and\ \citenamefont
  {Albert}(1999)}]{barabasi_emergence_1999}%
  \BibitemOpen
  \bibfield  {author} {\bibinfo {author} {\bibfnamefont {A.-L.}\ \bibnamefont
  {Barab{\'a}si}}\ and\ \bibinfo {author} {\bibfnamefont {R.}~\bibnamefont
  {Albert}},\ }\href {\doibase 10.1126/science.286.5439.509} {\bibfield
  {journal} {\bibinfo  {journal} {Science}\ }\textbf {\bibinfo {volume}
  {286}},\ \bibinfo {pages} {509} (\bibinfo {year} {1999})}\BibitemShut
  {NoStop}%
\bibitem [{\citenamefont {Barrat}\ and\ \citenamefont
  {Weigt}(2000)}]{barrat_properties_2000}%
  \BibitemOpen
  \bibfield  {author} {\bibinfo {author} {\bibfnamefont {A.}~\bibnamefont
  {Barrat}}\ and\ \bibinfo {author} {\bibfnamefont {M.}~\bibnamefont {Weigt}},\
  }\href {\doibase 10.1007/s100510050067} {\bibfield  {journal} {\bibinfo
  {journal} {Eur. Phys. J. B}\ }\textbf {\bibinfo {volume} {13}},\ \bibinfo
  {pages} {547} (\bibinfo {year} {2000})}\BibitemShut {NoStop}%
\bibitem [{\citenamefont {Strogatz}(2001)}]{strogatz_exploring_2001}%
  \BibitemOpen
  \bibfield  {author} {\bibinfo {author} {\bibfnamefont {S.~H.}\ \bibnamefont
  {Strogatz}},\ }\href {\doibase 10.1038/35065725} {\bibfield  {journal}
  {\bibinfo  {journal} {Nature}\ }\textbf {\bibinfo {volume} {410}},\ \bibinfo
  {pages} {268} (\bibinfo {year} {2001})}\BibitemShut {NoStop}%
\bibitem [{\citenamefont {Girvan}\ and\ \citenamefont
  {Newman}(2002)}]{girvan_community_2002}%
  \BibitemOpen
  \bibfield  {author} {\bibinfo {author} {\bibfnamefont {M.}~\bibnamefont
  {Girvan}}\ and\ \bibinfo {author} {\bibfnamefont {M.~E.~J.}\ \bibnamefont
  {Newman}},\ }\href {\doibase 10.1073/pnas.122653799} {\bibfield  {journal}
  {\bibinfo  {journal} {PNAS}\ }\textbf {\bibinfo {volume} {99}},\ \bibinfo
  {pages} {7821} (\bibinfo {year} {2002})}\BibitemShut {NoStop}%
\bibitem [{\citenamefont {Albert}\ and\ \citenamefont
  {Barab{\'a}si}(2002)}]{albert_statistical_2002}%
  \BibitemOpen
  \bibfield  {author} {\bibinfo {author} {\bibfnamefont {R.}~\bibnamefont
  {Albert}}\ and\ \bibinfo {author} {\bibfnamefont {A.-L.}\ \bibnamefont
  {Barab{\'a}si}},\ }\href {\doibase 10.1103/RevModPhys.74.47} {\bibfield
  {journal} {\bibinfo  {journal} {Rev. Mod. Phys.}\ }\textbf {\bibinfo {volume}
  {74}},\ \bibinfo {pages} {47} (\bibinfo {year} {2002})}\BibitemShut {NoStop}%
\bibitem [{\citenamefont {Barrat}\ \emph {et~al.}(2004)\citenamefont {Barrat},
  \citenamefont {Barth{\'e}lemy}, \citenamefont {Pastor-Satorras},\ and\
  \citenamefont {Vespignani}}]{barrat_architecture_2004}%
  \BibitemOpen
  \bibfield  {author} {\bibinfo {author} {\bibfnamefont {A.}~\bibnamefont
  {Barrat}}, \bibinfo {author} {\bibfnamefont {M.}~\bibnamefont
  {Barth{\'e}lemy}}, \bibinfo {author} {\bibfnamefont {R.}~\bibnamefont
  {Pastor-Satorras}}, \ and\ \bibinfo {author} {\bibfnamefont {A.}~\bibnamefont
  {Vespignani}},\ }\href {\doibase 10.1073/pnas.0400087101} {\bibfield
  {journal} {\bibinfo  {journal} {PNAS}\ }\textbf {\bibinfo {volume} {101}},\
  \bibinfo {pages} {3747} (\bibinfo {year} {2004})}\BibitemShut {NoStop}%
\bibitem [{\citenamefont {Dorogovtsev}\ \emph {et~al.}(2008)\citenamefont
  {Dorogovtsev}, \citenamefont {Goltsev},\ and\ \citenamefont
  {Mendes}}]{dorogovtsev_critical_2008}%
  \BibitemOpen
  \bibfield  {author} {\bibinfo {author} {\bibfnamefont {S.~N.}\ \bibnamefont
  {Dorogovtsev}}, \bibinfo {author} {\bibfnamefont {A.~V.}\ \bibnamefont
  {Goltsev}}, \ and\ \bibinfo {author} {\bibfnamefont {J.~F.~F.}\ \bibnamefont
  {Mendes}},\ }\href {\doibase 10.1103/RevModPhys.80.1275} {\bibfield
  {journal} {\bibinfo  {journal} {Rev. Mod. Phys.}\ }\textbf {\bibinfo {volume}
  {80}},\ \bibinfo {pages} {1275} (\bibinfo {year} {2008})}\BibitemShut
  {NoStop}%
\bibitem [{\citenamefont {Arenas}\ \emph {et~al.}(2008)\citenamefont {Arenas},
  \citenamefont {D{\'\i}az-Guilera}, \citenamefont {Kurths}, \citenamefont
  {Moreno},\ and\ \citenamefont {Zhou}}]{arenas_synchronization_2008}%
  \BibitemOpen
  \bibfield  {author} {\bibinfo {author} {\bibfnamefont {A.}~\bibnamefont
  {Arenas}}, \bibinfo {author} {\bibfnamefont {A.}~\bibnamefont
  {D{\'\i}az-Guilera}}, \bibinfo {author} {\bibfnamefont {J.}~\bibnamefont
  {Kurths}}, \bibinfo {author} {\bibfnamefont {Y.}~\bibnamefont {Moreno}}, \
  and\ \bibinfo {author} {\bibfnamefont {C.}~\bibnamefont {Zhou}},\ }\href
  {\doibase 10.1016/j.physrep.2008.09.002} {\bibfield  {journal} {\bibinfo
  {journal} {Physics Reports}\ }\textbf {\bibinfo {volume} {469}},\ \bibinfo
  {pages} {93} (\bibinfo {year} {2008})}\BibitemShut {NoStop}%
\bibitem [{\citenamefont {Pastor-Satorras}\ \emph {et~al.}(2015)\citenamefont
  {Pastor-Satorras}, \citenamefont {Castellano}, \citenamefont {Van~Mieghem},\
  and\ \citenamefont {Vespignani}}]{pastor-satorras_epidemic_2015}%
  \BibitemOpen
  \bibfield  {author} {\bibinfo {author} {\bibfnamefont {R.}~\bibnamefont
  {Pastor-Satorras}}, \bibinfo {author} {\bibfnamefont {C.}~\bibnamefont
  {Castellano}}, \bibinfo {author} {\bibfnamefont {P.}~\bibnamefont
  {Van~Mieghem}}, \ and\ \bibinfo {author} {\bibfnamefont {A.}~\bibnamefont
  {Vespignani}},\ }\href {\doibase 10.1103/RevModPhys.87.925} {\bibfield
  {journal} {\bibinfo  {journal} {Rev. Mod. Phys.}\ }\textbf {\bibinfo {volume}
  {87}},\ \bibinfo {pages} {925} (\bibinfo {year} {2015})}\BibitemShut
  {NoStop}%
\bibitem [{\citenamefont {Porter}(2012)}]{Porter:2012}%
  \BibitemOpen
  \bibfield  {author} {\bibinfo {author} {\bibfnamefont {M.~A.}\ \bibnamefont
  {Porter}},\ }\href {\doibase 10.4249/scholarpedia.1739} {\bibfield  {journal}
  {\bibinfo  {journal} {Scholarpedia}\ }\textbf {\bibinfo {volume} {7}},\
  \bibinfo {pages} {1739} (\bibinfo {year} {2012})}\BibitemShut {NoStop}%
\bibitem [{\citenamefont {Erd\"os}\ and\ \citenamefont
  {R\'enyi}(1959)}]{Erdos1959}%
  \BibitemOpen
  \bibfield  {author} {\bibinfo {author} {\bibfnamefont {P.}~\bibnamefont
  {Erd\"os}}\ and\ \bibinfo {author} {\bibfnamefont {A.}~\bibnamefont
  {R\'enyi}},\ }\href@noop {} {\bibfield  {journal} {\bibinfo  {journal}
  {Publicationes Mathematicae (Debrecen)}\ }\textbf {\bibinfo {volume} {6}},\
  \bibinfo {pages} {290} (\bibinfo {year} {1959})}\BibitemShut {NoStop}%
\bibitem [{\citenamefont {Monasson}(1999)}]{monasson_diffusion_1999}%
  \BibitemOpen
  \bibfield  {author} {\bibinfo {author} {\bibfnamefont {R.}~\bibnamefont
  {Monasson}},\ }\href {\doibase 10.1007/s100510051038} {\bibfield  {journal}
  {\bibinfo  {journal} {Eur. Phys. J. B}\ }\textbf {\bibinfo {volume} {12}},\
  \bibinfo {pages} {555} (\bibinfo {year} {1999})}\BibitemShut {NoStop}%
\bibitem [{\citenamefont {Newman}\ and\ \citenamefont
  {Watts}(1999)}]{newman_renormalization_1999}%
  \BibitemOpen
  \bibfield  {author} {\bibinfo {author} {\bibfnamefont {M.~E.~J.}\
  \bibnamefont {Newman}}\ and\ \bibinfo {author} {\bibfnamefont {D.~J.}\
  \bibnamefont {Watts}},\ }\href {\doibase 10.1016/S0375-9601(99)00757-4}
  {\bibfield  {journal} {\bibinfo  {journal} {Physics Letters A}\ }\textbf
  {\bibinfo {volume} {263}},\ \bibinfo {pages} {341} (\bibinfo {year}
  {1999})}\BibitemShut {NoStop}%
\bibitem [{\citenamefont {Hastings}(2003)}]{Hastings2003}%
  \BibitemOpen
  \bibfield  {author} {\bibinfo {author} {\bibfnamefont {M.~B.}\ \bibnamefont
  {Hastings}},\ }\href {\doibase 10.1103/PhysRevLett.91.098701} {\bibfield
  {journal} {\bibinfo  {journal} {Phys. Rev. Lett.}\ }\textbf {\bibinfo
  {volume} {91}},\ \bibinfo {pages} {098701} (\bibinfo {year}
  {2003})}\BibitemShut {NoStop}%
\bibitem [{\citenamefont {Barth{\'e}l{\'e}my}\ and\ \citenamefont
  {Amaral}(1999{\natexlab{a}})}]{barthelemy_small-world_1999}%
  \BibitemOpen
  \bibfield  {author} {\bibinfo {author} {\bibfnamefont {M.}~\bibnamefont
  {Barth{\'e}l{\'e}my}}\ and\ \bibinfo {author} {\bibfnamefont {L.~A.~N.}\
  \bibnamefont {Amaral}},\ }\href {\doibase 10.1103/PhysRevLett.82.3180}
  {\bibfield  {journal} {\bibinfo  {journal} {Phys. Rev. Lett.}\ }\textbf
  {\bibinfo {volume} {82}},\ \bibinfo {pages} {3180} (\bibinfo {year}
  {1999}{\natexlab{a}})}\BibitemShut {NoStop}%
\bibitem [{\citenamefont {Barth{\'e}l{\'e}my}\ and\ \citenamefont
  {Amaral}(1999{\natexlab{b}})}]{barthelemy_erratum_1999}%
  \BibitemOpen
  \bibfield  {author} {\bibinfo {author} {\bibfnamefont {M.}~\bibnamefont
  {Barth{\'e}l{\'e}my}}\ and\ \bibinfo {author} {\bibfnamefont {L.~A.~N.}\
  \bibnamefont {Amaral}},\ }\href {\doibase 10.1103/PhysRevLett.82.5180}
  {\bibfield  {journal} {\bibinfo  {journal} {Phys. Rev. Lett.}\ }\textbf
  {\bibinfo {volume} {82}},\ \bibinfo {pages} {5180} (\bibinfo {year}
  {1999}{\natexlab{b}})}\BibitemShut {NoStop}%
\bibitem [{\citenamefont {Gitterman}(2000)}]{gitterman_small-world_2000}%
  \BibitemOpen
  \bibfield  {author} {\bibinfo {author} {\bibfnamefont {M.}~\bibnamefont
  {Gitterman}},\ }\href {\doibase 10.1088/0305-4470/33/47/304} {\bibfield
  {journal} {\bibinfo  {journal} {J. Phys. A: Math. Gen.}\ }\textbf {\bibinfo
  {volume} {33}},\ \bibinfo {pages} {8373} (\bibinfo {year}
  {2000})}\BibitemShut {NoStop}%
\bibitem [{\citenamefont {Kim}\ \emph {et~al.}(2001)\citenamefont {Kim},
  \citenamefont {Hong}, \citenamefont {Holme}, \citenamefont {Jeon},
  \citenamefont {Minnhagen},\ and\ \citenamefont {Choi}}]{kim_xy_2001}%
  \BibitemOpen
  \bibfield  {author} {\bibinfo {author} {\bibfnamefont {B.~J.}\ \bibnamefont
  {Kim}}, \bibinfo {author} {\bibfnamefont {H.}~\bibnamefont {Hong}}, \bibinfo
  {author} {\bibfnamefont {P.}~\bibnamefont {Holme}}, \bibinfo {author}
  {\bibfnamefont {G.~S.}\ \bibnamefont {Jeon}}, \bibinfo {author}
  {\bibfnamefont {P.}~\bibnamefont {Minnhagen}}, \ and\ \bibinfo {author}
  {\bibfnamefont {M.~Y.}\ \bibnamefont {Choi}},\ }\href {\doibase
  10.1103/PhysRevE.64.056135} {\bibfield  {journal} {\bibinfo  {journal} {Phys.
  Rev. E}\ }\textbf {\bibinfo {volume} {64}},\ \bibinfo {pages} {056135}
  (\bibinfo {year} {2001})}\BibitemShut {NoStop}%
\bibitem [{\citenamefont {Hong}\ \emph {et~al.}(2002)\citenamefont {Hong},
  \citenamefont {Kim},\ and\ \citenamefont {Choi}}]{hong_comment_2002}%
  \BibitemOpen
  \bibfield  {author} {\bibinfo {author} {\bibfnamefont {H.}~\bibnamefont
  {Hong}}, \bibinfo {author} {\bibfnamefont {B.~J.}\ \bibnamefont {Kim}}, \
  and\ \bibinfo {author} {\bibfnamefont {M.~Y.}\ \bibnamefont {Choi}},\ }\href
  {\doibase 10.1103/PhysRevE.66.018101} {\bibfield  {journal} {\bibinfo
  {journal} {Phys. Rev. E}\ }\textbf {\bibinfo {volume} {66}},\ \bibinfo
  {pages} {018101} (\bibinfo {year} {2002})}\BibitemShut {NoStop}%
\bibitem [{\citenamefont {Herrero}(2002)}]{herrero_ising_2002}%
  \BibitemOpen
  \bibfield  {author} {\bibinfo {author} {\bibfnamefont {C.~P.}\ \bibnamefont
  {Herrero}},\ }\href {\doibase 10.1103/PhysRevE.65.066110} {\bibfield
  {journal} {\bibinfo  {journal} {Phys. Rev. E}\ }\textbf {\bibinfo {volume}
  {65}},\ \bibinfo {pages} {066110} (\bibinfo {year} {2002})}\BibitemShut
  {NoStop}%
\bibitem [{\citenamefont {Dorogovtsev}\ \emph {et~al.}(2002)\citenamefont
  {Dorogovtsev}, \citenamefont {Goltsev},\ and\ \citenamefont
  {Mendes}}]{dorogovtsev_ising_2002}%
  \BibitemOpen
  \bibfield  {author} {\bibinfo {author} {\bibfnamefont {S.~N.}\ \bibnamefont
  {Dorogovtsev}}, \bibinfo {author} {\bibfnamefont {A.~V.}\ \bibnamefont
  {Goltsev}}, \ and\ \bibinfo {author} {\bibfnamefont {J.~F.~F.}\ \bibnamefont
  {Mendes}},\ }\href {\doibase 10.1103/PhysRevE.66.016104} {\bibfield
  {journal} {\bibinfo  {journal} {Phys. Rev. E}\ }\textbf {\bibinfo {volume}
  {66}},\ \bibinfo {pages} {016104} (\bibinfo {year} {2002})}\BibitemShut
  {NoStop}%
\bibitem [{\citenamefont {Igl{\'o}i}\ and\ \citenamefont
  {Turban}(2002)}]{igloi_first_2002}%
  \BibitemOpen
  \bibfield  {author} {\bibinfo {author} {\bibfnamefont {F.}~\bibnamefont
  {Igl{\'o}i}}\ and\ \bibinfo {author} {\bibfnamefont {L.}~\bibnamefont
  {Turban}},\ }\href {\doibase 10.1103/PhysRevE.66.036140} {\bibfield
  {journal} {\bibinfo  {journal} {Phys. Rev. E}\ }\textbf {\bibinfo {volume}
  {66}},\ \bibinfo {pages} {036140} (\bibinfo {year} {2002})}\BibitemShut
  {NoStop}%
\bibitem [{\citenamefont {Goltsev}\ \emph {et~al.}(2003)\citenamefont
  {Goltsev}, \citenamefont {Dorogovtsev},\ and\ \citenamefont
  {Mendes}}]{goltsev_critical_2003}%
  \BibitemOpen
  \bibfield  {author} {\bibinfo {author} {\bibfnamefont {A.~V.}\ \bibnamefont
  {Goltsev}}, \bibinfo {author} {\bibfnamefont {S.~N.}\ \bibnamefont
  {Dorogovtsev}}, \ and\ \bibinfo {author} {\bibfnamefont {J.~F.~F.}\
  \bibnamefont {Mendes}},\ }\href {\doibase 10.1103/PhysRevE.67.026123}
  {\bibfield  {journal} {\bibinfo  {journal} {Phys. Rev. E}\ }\textbf {\bibinfo
  {volume} {67}},\ \bibinfo {pages} {026123} (\bibinfo {year}
  {2003})}\BibitemShut {NoStop}%
\bibitem [{\citenamefont {Herrero}(2004)}]{herrero_ising_2004}%
  \BibitemOpen
  \bibfield  {author} {\bibinfo {author} {\bibfnamefont {C.~P.}\ \bibnamefont
  {Herrero}},\ }\href {\doibase 10.1103/PhysRevE.69.067109} {\bibfield
  {journal} {\bibinfo  {journal} {Phys. Rev. E}\ }\textbf {\bibinfo {volume}
  {69}},\ \bibinfo {pages} {067109} (\bibinfo {year} {2004})}\BibitemShut
  {NoStop}%
\bibitem [{\citenamefont {Viana~Lopes}\ \emph {et~al.}(2004)\citenamefont
  {Viana~Lopes}, \citenamefont {Pogorelov}, \citenamefont {Lopes~dos Santos},\
  and\ \citenamefont {Toral}}]{lopes_exact_2004}%
  \BibitemOpen
  \bibfield  {author} {\bibinfo {author} {\bibfnamefont {J.}~\bibnamefont
  {Viana~Lopes}}, \bibinfo {author} {\bibfnamefont {Y.~G.}\ \bibnamefont
  {Pogorelov}}, \bibinfo {author} {\bibfnamefont {J.~M.~B.}\ \bibnamefont
  {Lopes~dos Santos}}, \ and\ \bibinfo {author} {\bibfnamefont
  {R.}~\bibnamefont {Toral}},\ }\href {\doibase 10.1103/PhysRevE.70.026112}
  {\bibfield  {journal} {\bibinfo  {journal} {Phys. Rev. E}\ }\textbf {\bibinfo
  {volume} {70}},\ \bibinfo {pages} {026112} (\bibinfo {year}
  {2004})}\BibitemShut {NoStop}%
\bibitem [{\citenamefont {Dorogovtsev}\ \emph {et~al.}(2004)\citenamefont
  {Dorogovtsev}, \citenamefont {Goltsev},\ and\ \citenamefont
  {Mendes}}]{dorogovtsev_potts_2004}%
  \BibitemOpen
  \bibfield  {author} {\bibinfo {author} {\bibfnamefont {S.~N.}\ \bibnamefont
  {Dorogovtsev}}, \bibinfo {author} {\bibfnamefont {A.~V.}\ \bibnamefont
  {Goltsev}}, \ and\ \bibinfo {author} {\bibfnamefont {J.~F.~F.}\ \bibnamefont
  {Mendes}},\ }\href {\doibase 10.1140/epjb/e2004-00019-y} {\bibfield
  {journal} {\bibinfo  {journal} {Eur. Phys. J. B}\ }\textbf {\bibinfo {volume}
  {38}},\ \bibinfo {pages} {177} (\bibinfo {year} {2004})}\BibitemShut
  {NoStop}%
\bibitem [{\citenamefont {Yi}\ and\ \citenamefont
  {Choi}(2003)}]{yi_effect_2003}%
  \BibitemOpen
  \bibfield  {author} {\bibinfo {author} {\bibfnamefont {H.}~\bibnamefont
  {Yi}}\ and\ \bibinfo {author} {\bibfnamefont {M.-S.}\ \bibnamefont {Choi}},\
  }\href {\doibase 10.1103/PhysRevE.67.056125} {\bibfield  {journal} {\bibinfo
  {journal} {Phys. Rev. E}\ }\textbf {\bibinfo {volume} {67}},\ \bibinfo
  {pages} {056125} (\bibinfo {year} {2003})}\BibitemShut {NoStop}%
\bibitem [{\citenamefont {Yi}(2010)}]{yi_quantum_2010}%
  \BibitemOpen
  \bibfield  {author} {\bibinfo {author} {\bibfnamefont {H.}~\bibnamefont
  {Yi}},\ }\href {\doibase 10.1103/PhysRevE.81.012103} {\bibfield  {journal}
  {\bibinfo  {journal} {Phys. Rev. E}\ }\textbf {\bibinfo {volume} {81}},\
  \bibinfo {pages} {012103} (\bibinfo {year} {2010})}\BibitemShut {NoStop}%
\bibitem [{\citenamefont {Baek}\ \emph {et~al.}(2011)\citenamefont {Baek},
  \citenamefont {Um}, \citenamefont {Yi},\ and\ \citenamefont
  {Kim}}]{baek_quantum_2011}%
  \BibitemOpen
  \bibfield  {author} {\bibinfo {author} {\bibfnamefont {S.~K.}\ \bibnamefont
  {Baek}}, \bibinfo {author} {\bibfnamefont {J.}~\bibnamefont {Um}}, \bibinfo
  {author} {\bibfnamefont {S.~D.}\ \bibnamefont {Yi}}, \ and\ \bibinfo {author}
  {\bibfnamefont {B.~J.}\ \bibnamefont {Kim}},\ }\href {\doibase
  10.1103/PhysRevB.84.174419} {\bibfield  {journal} {\bibinfo  {journal} {Phys.
  Rev. B}\ }\textbf {\bibinfo {volume} {84}},\ \bibinfo {pages} {174419}
  (\bibinfo {year} {2011})}\BibitemShut {NoStop}%
\bibitem [{\citenamefont {Yi}(2015)}]{yi_quantum_2015}%
  \BibitemOpen
  \bibfield  {author} {\bibinfo {author} {\bibfnamefont {H.}~\bibnamefont
  {Yi}},\ }\href {\doibase 10.1103/PhysRevE.91.012146} {\bibfield  {journal}
  {\bibinfo  {journal} {Phys. Rev. E}\ }\textbf {\bibinfo {volume} {91}},\
  \bibinfo {pages} {012146} (\bibinfo {year} {2015})}\BibitemShut {NoStop}%
\bibitem [{\citenamefont {Medvedyeva}\ \emph {et~al.}(2003)\citenamefont
  {Medvedyeva}, \citenamefont {Holme}, \citenamefont {Minnhagen},\ and\
  \citenamefont {Kim}}]{medvedyeva_dynamic_2003}%
  \BibitemOpen
  \bibfield  {author} {\bibinfo {author} {\bibfnamefont {K.}~\bibnamefont
  {Medvedyeva}}, \bibinfo {author} {\bibfnamefont {P.}~\bibnamefont {Holme}},
  \bibinfo {author} {\bibfnamefont {P.}~\bibnamefont {Minnhagen}}, \ and\
  \bibinfo {author} {\bibfnamefont {B.~J.}\ \bibnamefont {Kim}},\ }\href
  {\doibase 10.1103/PhysRevE.67.036118} {\bibfield  {journal} {\bibinfo
  {journal} {Phys. Rev. E}\ }\textbf {\bibinfo {volume} {67}},\ \bibinfo
  {pages} {036118} (\bibinfo {year} {2003})}\BibitemShut {NoStop}%
\bibitem [{\citenamefont {De~Nigris}\ and\ \citenamefont
  {Leoncini}(2013)}]{de_nigris_critical_2013}%
  \BibitemOpen
  \bibfield  {author} {\bibinfo {author} {\bibfnamefont {S.}~\bibnamefont
  {De~Nigris}}\ and\ \bibinfo {author} {\bibfnamefont {X.}~\bibnamefont
  {Leoncini}},\ }\href {\doibase 10.1103/PhysRevE.88.012131} {\bibfield
  {journal} {\bibinfo  {journal} {Phys. Rev. E}\ }\textbf {\bibinfo {volume}
  {88}},\ \bibinfo {pages} {012131} (\bibinfo {year} {2013})}\BibitemShut
  {NoStop}%
\bibitem [{\citenamefont {Jeong}\ \emph {et~al.}(2003)\citenamefont {Jeong},
  \citenamefont {Hong}, \citenamefont {Kim},\ and\ \citenamefont
  {Choi}}]{jeong_phase_2003}%
  \BibitemOpen
  \bibfield  {author} {\bibinfo {author} {\bibfnamefont {D.}~\bibnamefont
  {Jeong}}, \bibinfo {author} {\bibfnamefont {H.}~\bibnamefont {Hong}},
  \bibinfo {author} {\bibfnamefont {B.~J.}\ \bibnamefont {Kim}}, \ and\
  \bibinfo {author} {\bibfnamefont {M.~Y.}\ \bibnamefont {Choi}},\ }\href
  {\doibase 10.1103/PhysRevE.68.027101} {\bibfield  {journal} {\bibinfo
  {journal} {Phys. Rev. E}\ }\textbf {\bibinfo {volume} {68}},\ \bibinfo
  {pages} {027101} (\bibinfo {year} {2003})}\BibitemShut {NoStop}%
\bibitem [{\citenamefont {Chatterjee}\ and\ \citenamefont
  {Sen}(2006)}]{chatterjee_phase_2006}%
  \BibitemOpen
  \bibfield  {author} {\bibinfo {author} {\bibfnamefont {A.}~\bibnamefont
  {Chatterjee}}\ and\ \bibinfo {author} {\bibfnamefont {P.}~\bibnamefont
  {Sen}},\ }\href {\doibase 10.1103/PhysRevE.74.036109} {\bibfield  {journal}
  {\bibinfo  {journal} {Phys. Rev. E}\ }\textbf {\bibinfo {volume} {74}},\
  \bibinfo {pages} {036109} (\bibinfo {year} {2006})}\BibitemShut {NoStop}%
\bibitem [{\citenamefont {Fisher}\ \emph {et~al.}(1972)\citenamefont {Fisher},
  \citenamefont {Ma},\ and\ \citenamefont {Nickel}}]{fisher_critical_1972}%
  \BibitemOpen
  \bibfield  {author} {\bibinfo {author} {\bibfnamefont {M.~E.}\ \bibnamefont
  {Fisher}}, \bibinfo {author} {\bibfnamefont {S.-k.}\ \bibnamefont {Ma}}, \
  and\ \bibinfo {author} {\bibfnamefont {B.~G.}\ \bibnamefont {Nickel}},\
  }\href {\doibase 10.1103/PhysRevLett.29.917} {\bibfield  {journal} {\bibinfo
  {journal} {Phys. Rev. Lett.}\ }\textbf {\bibinfo {volume} {29}},\ \bibinfo
  {pages} {917} (\bibinfo {year} {1972})}\BibitemShut {NoStop}%
\bibitem [{\citenamefont {Brezin}\ \emph {et~al.}(1976)\citenamefont {Brezin},
  \citenamefont {Zinn-Justin},\ and\ \citenamefont
  {Guillou}}]{brezin_critical_1976}%
  \BibitemOpen
  \bibfield  {author} {\bibinfo {author} {\bibfnamefont {E.}~\bibnamefont
  {Brezin}}, \bibinfo {author} {\bibfnamefont {J.}~\bibnamefont {Zinn-Justin}},
  \ and\ \bibinfo {author} {\bibfnamefont {J.~C.~L.}\ \bibnamefont {Guillou}},\
  }\href {\doibase 10.1088/0305-4470/9/9/003} {\bibfield  {journal} {\bibinfo
  {journal} {J. Phys. A: Math. Gen.}\ }\textbf {\bibinfo {volume} {9}},\
  \bibinfo {pages} {L119} (\bibinfo {year} {1976})}\BibitemShut {NoStop}%
\bibitem [{\citenamefont {Yusuf}\ \emph {et~al.}(2004)\citenamefont {Yusuf},
  \citenamefont {Joshi},\ and\ \citenamefont {Yang}}]{PhysRevB.69.144412}%
  \BibitemOpen
  \bibfield  {author} {\bibinfo {author} {\bibfnamefont {E.}~\bibnamefont
  {Yusuf}}, \bibinfo {author} {\bibfnamefont {A.}~\bibnamefont {Joshi}}, \ and\
  \bibinfo {author} {\bibfnamefont {K.}~\bibnamefont {Yang}},\ }\href {\doibase
  10.1103/PhysRevB.69.144412} {\bibfield  {journal} {\bibinfo  {journal} {Phys.
  Rev. B}\ }\textbf {\bibinfo {volume} {69}},\ \bibinfo {pages} {144412}
  (\bibinfo {year} {2004})}\BibitemShut {NoStop}%
\bibitem [{\citenamefont {Laflorencie}\ \emph {et~al.}(2005)\citenamefont
  {Laflorencie}, \citenamefont {Affleck},\ and\ \citenamefont
  {Berciu}}]{Laflorencie_2005}%
  \BibitemOpen
  \bibfield  {author} {\bibinfo {author} {\bibfnamefont {N.}~\bibnamefont
  {Laflorencie}}, \bibinfo {author} {\bibfnamefont {I.}~\bibnamefont
  {Affleck}}, \ and\ \bibinfo {author} {\bibfnamefont {M.}~\bibnamefont
  {Berciu}},\ }\href {\doibase 10.1088/1742-5468/2005/12/p12001} {\bibfield
  {journal} {\bibinfo  {journal} {Journal of Statistical Mechanics: Theory and
  Experiment}\ }\textbf {\bibinfo {volume} {2005}},\ \bibinfo {pages} {P12001}
  (\bibinfo {year} {2005})}\BibitemShut {NoStop}%
\bibitem [{\citenamefont {Kardar}(2007)}]{kardar_2007}%
  \BibitemOpen
  \bibfield  {author} {\bibinfo {author} {\bibfnamefont {M.}~\bibnamefont
  {Kardar}},\ }\href {\doibase 10.1017/CBO9780511815881} {\emph {\bibinfo
  {title} {Statistical Physics of Fields}}}\ (\bibinfo  {publisher} {Cambridge
  University Press},\ \bibinfo {year} {2007})\BibitemShut {NoStop}%
\bibitem [{\citenamefont {Scalapino}\ \emph {et~al.}(1975)\citenamefont
  {Scalapino}, \citenamefont {Imry},\ and\ \citenamefont
  {Pincus}}]{Scalapino1975}%
  \BibitemOpen
  \bibfield  {author} {\bibinfo {author} {\bibfnamefont {D.~J.}\ \bibnamefont
  {Scalapino}}, \bibinfo {author} {\bibfnamefont {Y.}~\bibnamefont {Imry}}, \
  and\ \bibinfo {author} {\bibfnamefont {P.}~\bibnamefont {Pincus}},\ }\href
  {\doibase 10.1103/PhysRevB.11.2042} {\bibfield  {journal} {\bibinfo
  {journal} {Phys. Rev. B}\ }\textbf {\bibinfo {volume} {11}},\ \bibinfo
  {pages} {2042} (\bibinfo {year} {1975})}\BibitemShut {NoStop}%
\bibitem [{\citenamefont {Schulz}(1996)}]{Schulz1996}%
  \BibitemOpen
  \bibfield  {author} {\bibinfo {author} {\bibfnamefont {H.~J.}\ \bibnamefont
  {Schulz}},\ }\href {\doibase 10.1103/PhysRevLett.77.2790} {\bibfield
  {journal} {\bibinfo  {journal} {Phys. Rev. Lett.}\ }\textbf {\bibinfo
  {volume} {77}},\ \bibinfo {pages} {2790} (\bibinfo {year}
  {1996})}\BibitemShut {NoStop}%
\bibitem [{\citenamefont {Yasuda}\ \emph {et~al.}(2005)\citenamefont {Yasuda},
  \citenamefont {Todo}, \citenamefont {Hukushima}, \citenamefont {Alet},
  \citenamefont {Keller}, \citenamefont {Troyer},\ and\ \citenamefont
  {Takayama}}]{Yasuda2005}%
  \BibitemOpen
  \bibfield  {author} {\bibinfo {author} {\bibfnamefont {C.}~\bibnamefont
  {Yasuda}}, \bibinfo {author} {\bibfnamefont {S.}~\bibnamefont {Todo}},
  \bibinfo {author} {\bibfnamefont {K.}~\bibnamefont {Hukushima}}, \bibinfo
  {author} {\bibfnamefont {F.}~\bibnamefont {Alet}}, \bibinfo {author}
  {\bibfnamefont {M.}~\bibnamefont {Keller}}, \bibinfo {author} {\bibfnamefont
  {M.}~\bibnamefont {Troyer}}, \ and\ \bibinfo {author} {\bibfnamefont
  {H.}~\bibnamefont {Takayama}},\ }\href {\doibase
  10.1103/PhysRevLett.94.217201} {\bibfield  {journal} {\bibinfo  {journal}
  {Phys. Rev. Lett.}\ }\textbf {\bibinfo {volume} {94}},\ \bibinfo {pages}
  {217201} (\bibinfo {year} {2005})}\BibitemShut {NoStop}%
\bibitem [{\citenamefont {Thielemann}\ \emph {et~al.}(2009)\citenamefont
  {Thielemann}, \citenamefont {R{\"u}egg}, \citenamefont {Kiefer},
  \citenamefont {R{\o}nnow}, \citenamefont {Normand}, \citenamefont {Bouillot},
  \citenamefont {Kollath}, \citenamefont {Orignac}, \citenamefont {Citro},
  \citenamefont {Giamarchi}, \citenamefont {L{\"a}uchli}, \citenamefont
  {Biner}, \citenamefont {Kr{\"a}mer}, \citenamefont {Wolff-Fabris},
  \citenamefont {Zapf}, \citenamefont {Jaime}, \citenamefont {Stahn},
  \citenamefont {Christensen}, \citenamefont {Grenier}, \citenamefont
  {McMorrow},\ and\ \citenamefont {Mesot}}]{thielemann_field-controlled_2009}%
  \BibitemOpen
  \bibfield  {author} {\bibinfo {author} {\bibfnamefont {B.}~\bibnamefont
  {Thielemann}}, \bibinfo {author} {\bibfnamefont {C.}~\bibnamefont
  {R{\"u}egg}}, \bibinfo {author} {\bibfnamefont {K.}~\bibnamefont {Kiefer}},
  \bibinfo {author} {\bibfnamefont {H.~M.}\ \bibnamefont {R{\o}nnow}}, \bibinfo
  {author} {\bibfnamefont {B.}~\bibnamefont {Normand}}, \bibinfo {author}
  {\bibfnamefont {P.}~\bibnamefont {Bouillot}}, \bibinfo {author}
  {\bibfnamefont {C.}~\bibnamefont {Kollath}}, \bibinfo {author} {\bibfnamefont
  {E.}~\bibnamefont {Orignac}}, \bibinfo {author} {\bibfnamefont
  {R.}~\bibnamefont {Citro}}, \bibinfo {author} {\bibfnamefont
  {T.}~\bibnamefont {Giamarchi}}, \bibinfo {author} {\bibfnamefont {A.~M.}\
  \bibnamefont {L{\"a}uchli}}, \bibinfo {author} {\bibfnamefont
  {D.}~\bibnamefont {Biner}}, \bibinfo {author} {\bibfnamefont {K.~W.}\
  \bibnamefont {Kr{\"a}mer}}, \bibinfo {author} {\bibfnamefont
  {F.}~\bibnamefont {Wolff-Fabris}}, \bibinfo {author} {\bibfnamefont {V.~S.}\
  \bibnamefont {Zapf}}, \bibinfo {author} {\bibfnamefont {M.}~\bibnamefont
  {Jaime}}, \bibinfo {author} {\bibfnamefont {J.}~\bibnamefont {Stahn}},
  \bibinfo {author} {\bibfnamefont {N.~B.}\ \bibnamefont {Christensen}},
  \bibinfo {author} {\bibfnamefont {B.}~\bibnamefont {Grenier}}, \bibinfo
  {author} {\bibfnamefont {D.~F.}\ \bibnamefont {McMorrow}}, \ and\ \bibinfo
  {author} {\bibfnamefont {J.}~\bibnamefont {Mesot}},\ }\href {\doibase
  10.1103/PhysRevB.79.020408} {\bibfield  {journal} {\bibinfo  {journal} {Phys.
  Rev. B}\ }\textbf {\bibinfo {volume} {79}},\ \bibinfo {pages} {020408}
  (\bibinfo {year} {2009})}\BibitemShut {NoStop}%
\bibitem [{\citenamefont {Bouillot}\ \emph {et~al.}(2011)\citenamefont
  {Bouillot}, \citenamefont {Kollath}, \citenamefont {L{\"a}uchli},
  \citenamefont {Zvonarev}, \citenamefont {Thielemann}, \citenamefont
  {R{\"u}egg}, \citenamefont {Orignac}, \citenamefont {Citro}, \citenamefont
  {Klanj{\v s}ek}, \citenamefont {Berthier}, \citenamefont {Horvati{\'c}},\
  and\ \citenamefont {Giamarchi}}]{bouillot_statics_2011}%
  \BibitemOpen
  \bibfield  {author} {\bibinfo {author} {\bibfnamefont {P.}~\bibnamefont
  {Bouillot}}, \bibinfo {author} {\bibfnamefont {C.}~\bibnamefont {Kollath}},
  \bibinfo {author} {\bibfnamefont {A.~M.}\ \bibnamefont {L{\"a}uchli}},
  \bibinfo {author} {\bibfnamefont {M.}~\bibnamefont {Zvonarev}}, \bibinfo
  {author} {\bibfnamefont {B.}~\bibnamefont {Thielemann}}, \bibinfo {author}
  {\bibfnamefont {C.}~\bibnamefont {R{\"u}egg}}, \bibinfo {author}
  {\bibfnamefont {E.}~\bibnamefont {Orignac}}, \bibinfo {author} {\bibfnamefont
  {R.}~\bibnamefont {Citro}}, \bibinfo {author} {\bibfnamefont
  {M.}~\bibnamefont {Klanj{\v s}ek}}, \bibinfo {author} {\bibfnamefont
  {C.}~\bibnamefont {Berthier}}, \bibinfo {author} {\bibfnamefont
  {M.}~\bibnamefont {Horvati{\'c}}}, \ and\ \bibinfo {author} {\bibfnamefont
  {T.}~\bibnamefont {Giamarchi}},\ }\href {\doibase 10.1103/PhysRevB.83.054407}
  {\bibfield  {journal} {\bibinfo  {journal} {Phys. Rev. B}\ }\textbf {\bibinfo
  {volume} {83}},\ \bibinfo {pages} {054407} (\bibinfo {year}
  {2011})}\BibitemShut {NoStop}%
\bibitem [{\citenamefont {Blinder}\ \emph {et~al.}(2017)\citenamefont
  {Blinder}, \citenamefont {Dupont}, \citenamefont {Mukhopadhyay},
  \citenamefont {Grbi{\'c}}, \citenamefont {Laflorencie}, \citenamefont
  {Capponi}, \citenamefont {Mayaffre}, \citenamefont {Berthier}, \citenamefont
  {Paduan-Filho},\ and\ \citenamefont {Horvati{\'c}}}]{blinder_nuclear_2017}%
  \BibitemOpen
  \bibfield  {author} {\bibinfo {author} {\bibfnamefont {R.}~\bibnamefont
  {Blinder}}, \bibinfo {author} {\bibfnamefont {M.}~\bibnamefont {Dupont}},
  \bibinfo {author} {\bibfnamefont {S.}~\bibnamefont {Mukhopadhyay}}, \bibinfo
  {author} {\bibfnamefont {M.~S.}\ \bibnamefont {Grbi{\'c}}}, \bibinfo {author}
  {\bibfnamefont {N.}~\bibnamefont {Laflorencie}}, \bibinfo {author}
  {\bibfnamefont {S.}~\bibnamefont {Capponi}}, \bibinfo {author} {\bibfnamefont
  {H.}~\bibnamefont {Mayaffre}}, \bibinfo {author} {\bibfnamefont
  {C.}~\bibnamefont {Berthier}}, \bibinfo {author} {\bibfnamefont
  {A.}~\bibnamefont {Paduan-Filho}}, \ and\ \bibinfo {author} {\bibfnamefont
  {M.}~\bibnamefont {Horvati{\'c}}},\ }\href {\doibase
  10.1103/PhysRevB.95.020404} {\bibfield  {journal} {\bibinfo  {journal} {Phys.
  Rev. B}\ }\textbf {\bibinfo {volume} {95}},\ \bibinfo {pages} {020404}
  (\bibinfo {year} {2017})}\BibitemShut {NoStop}%
\bibitem [{\citenamefont {Yao}\ and\ \citenamefont
  {Sandvik}(2007)}]{yao_universal_2007}%
  \BibitemOpen
  \bibfield  {author} {\bibinfo {author} {\bibfnamefont {D.~X.}\ \bibnamefont
  {Yao}}\ and\ \bibinfo {author} {\bibfnamefont {A.~W.}\ \bibnamefont
  {Sandvik}},\ }\href {\doibase 10.1103/PhysRevB.75.052411} {\bibfield
  {journal} {\bibinfo  {journal} {Phys. Rev. B}\ }\textbf {\bibinfo {volume}
  {75}},\ \bibinfo {pages} {052411} (\bibinfo {year} {2007})}\BibitemShut
  {NoStop}%
\bibitem [{\citenamefont {Lancaster}\ \emph {et~al.}(2007)\citenamefont
  {Lancaster}, \citenamefont {Blundell}, \citenamefont {Brooks}, \citenamefont
  {Baker}, \citenamefont {Pratt}, \citenamefont {Manson}, \citenamefont
  {Conner}, \citenamefont {Xiao}, \citenamefont {Landee}, \citenamefont
  {Chaves}, \citenamefont {Soriano}, \citenamefont {Novak}, \citenamefont
  {Papageorgiou}, \citenamefont {Bianchi}, \citenamefont {Herrmannsd{\"o}rfer},
  \citenamefont {Wosnitza},\ and\ \citenamefont
  {Schlueter}}]{lancaster_magnetic_2007}%
  \BibitemOpen
  \bibfield  {author} {\bibinfo {author} {\bibfnamefont {T.}~\bibnamefont
  {Lancaster}}, \bibinfo {author} {\bibfnamefont {S.~J.}\ \bibnamefont
  {Blundell}}, \bibinfo {author} {\bibfnamefont {M.~L.}\ \bibnamefont
  {Brooks}}, \bibinfo {author} {\bibfnamefont {P.~J.}\ \bibnamefont {Baker}},
  \bibinfo {author} {\bibfnamefont {F.~L.}\ \bibnamefont {Pratt}}, \bibinfo
  {author} {\bibfnamefont {J.~L.}\ \bibnamefont {Manson}}, \bibinfo {author}
  {\bibfnamefont {M.~M.}\ \bibnamefont {Conner}}, \bibinfo {author}
  {\bibfnamefont {F.}~\bibnamefont {Xiao}}, \bibinfo {author} {\bibfnamefont
  {C.~P.}\ \bibnamefont {Landee}}, \bibinfo {author} {\bibfnamefont {F.~A.}\
  \bibnamefont {Chaves}}, \bibinfo {author} {\bibfnamefont {S.}~\bibnamefont
  {Soriano}}, \bibinfo {author} {\bibfnamefont {M.~A.}\ \bibnamefont {Novak}},
  \bibinfo {author} {\bibfnamefont {T.~P.}\ \bibnamefont {Papageorgiou}},
  \bibinfo {author} {\bibfnamefont {A.~D.}\ \bibnamefont {Bianchi}}, \bibinfo
  {author} {\bibfnamefont {T.}~\bibnamefont {Herrmannsd{\"o}rfer}}, \bibinfo
  {author} {\bibfnamefont {J.}~\bibnamefont {Wosnitza}}, \ and\ \bibinfo
  {author} {\bibfnamefont {J.~A.}\ \bibnamefont {Schlueter}},\ }\href {\doibase
  10.1103/PhysRevB.75.094421} {\bibfield  {journal} {\bibinfo  {journal} {Phys.
  Rev. B}\ }\textbf {\bibinfo {volume} {75}},\ \bibinfo {pages} {094421}
  (\bibinfo {year} {2007})}\BibitemShut {NoStop}%
\bibitem [{\citenamefont {Goddard}\ \emph {et~al.}(2008)\citenamefont
  {Goddard}, \citenamefont {Singleton}, \citenamefont {Sengupta}, \citenamefont
  {McDonald}, \citenamefont {Lancaster}, \citenamefont {Blundell},
  \citenamefont {Pratt}, \citenamefont {Cox}, \citenamefont {Harrison},
  \citenamefont {Manson}, \citenamefont {Southerland},\ and\ \citenamefont
  {Schlueter}}]{goddard_experimentally_2008}%
  \BibitemOpen
  \bibfield  {author} {\bibinfo {author} {\bibfnamefont {P.~A.}\ \bibnamefont
  {Goddard}}, \bibinfo {author} {\bibfnamefont {J.}~\bibnamefont {Singleton}},
  \bibinfo {author} {\bibfnamefont {P.}~\bibnamefont {Sengupta}}, \bibinfo
  {author} {\bibfnamefont {R.~D.}\ \bibnamefont {McDonald}}, \bibinfo {author}
  {\bibfnamefont {T.}~\bibnamefont {Lancaster}}, \bibinfo {author}
  {\bibfnamefont {S.~J.}\ \bibnamefont {Blundell}}, \bibinfo {author}
  {\bibfnamefont {F.~L.}\ \bibnamefont {Pratt}}, \bibinfo {author}
  {\bibfnamefont {S.}~\bibnamefont {Cox}}, \bibinfo {author} {\bibfnamefont
  {N.}~\bibnamefont {Harrison}}, \bibinfo {author} {\bibfnamefont {J.~L.}\
  \bibnamefont {Manson}}, \bibinfo {author} {\bibfnamefont {H.~I.}\
  \bibnamefont {Southerland}}, \ and\ \bibinfo {author} {\bibfnamefont {J.~A.}\
  \bibnamefont {Schlueter}},\ }\href {\doibase 10.1088/1367-2630/10/8/083025}
  {\bibfield  {journal} {\bibinfo  {journal} {New J. Phys.}\ }\textbf {\bibinfo
  {volume} {10}},\ \bibinfo {pages} {083025} (\bibinfo {year}
  {2008})}\BibitemShut {NoStop}%
\bibitem [{\citenamefont {Juh{\'a}sz~Junger}\ \emph {et~al.}(2009)\citenamefont
  {Juh{\'a}sz~Junger}, \citenamefont {Ihle},\ and\ \citenamefont
  {Richter}}]{juhasz_junger_thermodynamics_2009}%
  \BibitemOpen
  \bibfield  {author} {\bibinfo {author} {\bibfnamefont {I.}~\bibnamefont
  {Juh{\'a}sz~Junger}}, \bibinfo {author} {\bibfnamefont {D.}~\bibnamefont
  {Ihle}}, \ and\ \bibinfo {author} {\bibfnamefont {J.}~\bibnamefont
  {Richter}},\ }\href {\doibase 10.1103/PhysRevB.80.064425} {\bibfield
  {journal} {\bibinfo  {journal} {Phys. Rev. B}\ }\textbf {\bibinfo {volume}
  {80}},\ \bibinfo {pages} {064425} (\bibinfo {year} {2009})}\BibitemShut
  {NoStop}%
\bibitem [{\citenamefont {Johnston}(2011)}]{johnston_magnetic_2011}%
  \BibitemOpen
  \bibfield  {author} {\bibinfo {author} {\bibfnamefont {D.~C.}\ \bibnamefont
  {Johnston}},\ }\href {\doibase 10.1103/PhysRevB.84.094445} {\bibfield
  {journal} {\bibinfo  {journal} {Phys. Rev. B}\ }\textbf {\bibinfo {volume}
  {84}} (\bibinfo {year} {2011}),\ 10.1103/PhysRevB.84.094445}\BibitemShut
  {NoStop}%
\bibitem [{\citenamefont {Gibertini}\ \emph {et~al.}(2019)\citenamefont
  {Gibertini}, \citenamefont {Koperski}, \citenamefont {Morpurgo},\ and\
  \citenamefont {Novoselov}}]{gibertini_magnetic_2019}%
  \BibitemOpen
  \bibfield  {author} {\bibinfo {author} {\bibfnamefont {M.}~\bibnamefont
  {Gibertini}}, \bibinfo {author} {\bibfnamefont {M.}~\bibnamefont {Koperski}},
  \bibinfo {author} {\bibfnamefont {A.~F.}\ \bibnamefont {Morpurgo}}, \ and\
  \bibinfo {author} {\bibfnamefont {K.~S.}\ \bibnamefont {Novoselov}},\ }\href
  {\doibase 10.1038/s41565-019-0438-6} {\bibfield  {journal} {\bibinfo
  {journal} {Nature Nanotechnology}\ }\textbf {\bibinfo {volume} {14}},\
  \bibinfo {pages} {408} (\bibinfo {year} {2019})}\BibitemShut {NoStop}%
\bibitem [{\citenamefont {Bollmark}\ \emph {et~al.}(2020)\citenamefont
  {Bollmark}, \citenamefont {Laflorencie},\ and\ \citenamefont
  {Kantian}}]{bollmark_dimensional_2020}%
  \BibitemOpen
  \bibfield  {author} {\bibinfo {author} {\bibfnamefont {G.}~\bibnamefont
  {Bollmark}}, \bibinfo {author} {\bibfnamefont {N.}~\bibnamefont
  {Laflorencie}}, \ and\ \bibinfo {author} {\bibfnamefont {A.}~\bibnamefont
  {Kantian}},\ }\href {\doibase 10.1103/PhysRevB.102.195145} {\bibfield
  {journal} {\bibinfo  {journal} {Phys. Rev. B}\ }\textbf {\bibinfo {volume}
  {102}},\ \bibinfo {pages} {195145} (\bibinfo {year} {2020})}\BibitemShut
  {NoStop}%
\bibitem [{\citenamefont {Irkhin}\ and\ \citenamefont
  {Katanin}(2000)}]{PhysRevB.61.6757}%
  \BibitemOpen
  \bibfield  {author} {\bibinfo {author} {\bibfnamefont {V.~Y.}\ \bibnamefont
  {Irkhin}}\ and\ \bibinfo {author} {\bibfnamefont {A.~A.}\ \bibnamefont
  {Katanin}},\ }\href {\doibase 10.1103/PhysRevB.61.6757} {\bibfield  {journal}
  {\bibinfo  {journal} {Phys. Rev. B}\ }\textbf {\bibinfo {volume} {61}},\
  \bibinfo {pages} {6757} (\bibinfo {year} {2000})}\BibitemShut {NoStop}%
\bibitem [{\citenamefont {Praz}\ \emph {et~al.}(2006)\citenamefont {Praz},
  \citenamefont {Mudry},\ and\ \citenamefont {Hastings}}]{PhysRevB.74.184407}%
  \BibitemOpen
  \bibfield  {author} {\bibinfo {author} {\bibfnamefont {A.}~\bibnamefont
  {Praz}}, \bibinfo {author} {\bibfnamefont {C.}~\bibnamefont {Mudry}}, \ and\
  \bibinfo {author} {\bibfnamefont {M.~B.}\ \bibnamefont {Hastings}},\ }\href
  {\doibase 10.1103/PhysRevB.74.184407} {\bibfield  {journal} {\bibinfo
  {journal} {Phys. Rev. B}\ }\textbf {\bibinfo {volume} {74}},\ \bibinfo
  {pages} {184407} (\bibinfo {year} {2006})}\BibitemShut {NoStop}%
\bibitem [{\citenamefont {Giamarchi}(2003)}]{giamarchi2003quantum}%
  \BibitemOpen
  \bibfield  {author} {\bibinfo {author} {\bibfnamefont {T.}~\bibnamefont
  {Giamarchi}},\ }\href@noop {} {\emph {\bibinfo {title} {Quantum physics in
  one dimension}}},\ Vol.\ \bibinfo {volume} {121}\ (\bibinfo  {publisher}
  {Clarendon press, Oxford},\ \bibinfo {year} {2003})\BibitemShut {NoStop}%
\bibitem [{\citenamefont {Korepin}\ \emph {et~al.}(1993)\citenamefont
  {Korepin}, \citenamefont {Bogoliubov},\ and\ \citenamefont
  {Izergin}}]{korepin_bogoliubov_izergin_1993}%
  \BibitemOpen
  \bibfield  {author} {\bibinfo {author} {\bibfnamefont {V.~E.}\ \bibnamefont
  {Korepin}}, \bibinfo {author} {\bibfnamefont {N.~M.}\ \bibnamefont
  {Bogoliubov}}, \ and\ \bibinfo {author} {\bibfnamefont {A.~G.}\ \bibnamefont
  {Izergin}},\ }\href {\doibase 10.1017/CBO9780511628832} {\emph {\bibinfo
  {title} {Quantum Inverse Scattering Method and Correlation Functions}}},\
  Cambridge Monographs on Mathematical Physics\ (\bibinfo  {publisher}
  {Cambridge University Press},\ \bibinfo {year} {1993})\BibitemShut {NoStop}%
\bibitem [{\citenamefont {Lukyanov}\ and\ \citenamefont
  {Zamolodchikov}(1997)}]{LUKYANOV1997571}%
  \BibitemOpen
  \bibfield  {author} {\bibinfo {author} {\bibfnamefont {S.}~\bibnamefont
  {Lukyanov}}\ and\ \bibinfo {author} {\bibfnamefont {A.}~\bibnamefont
  {Zamolodchikov}},\ }\href {\doibase
  https://doi.org/10.1016/S0550-3213(97)00123-5} {\bibfield  {journal}
  {\bibinfo  {journal} {Nuclear Physics B}\ }\textbf {\bibinfo {volume}
  {493}},\ \bibinfo {pages} {571 } (\bibinfo {year} {1997})}\BibitemShut
  {NoStop}%
\bibitem [{\citenamefont {Affleck}\ \emph {et~al.}(1989)\citenamefont
  {Affleck}, \citenamefont {Gepner}, \citenamefont {Schulz},\ and\
  \citenamefont {Ziman}}]{Affleck_1989}%
  \BibitemOpen
  \bibfield  {author} {\bibinfo {author} {\bibfnamefont {I.}~\bibnamefont
  {Affleck}}, \bibinfo {author} {\bibfnamefont {D.}~\bibnamefont {Gepner}},
  \bibinfo {author} {\bibfnamefont {H.~J.}\ \bibnamefont {Schulz}}, \ and\
  \bibinfo {author} {\bibfnamefont {T.}~\bibnamefont {Ziman}},\ }\href
  {\doibase 10.1088/0305-4470/22/5/015} {\bibfield  {journal} {\bibinfo
  {journal} {Journal of Physics A: Mathematical and General}\ }\textbf
  {\bibinfo {volume} {22}},\ \bibinfo {pages} {511} (\bibinfo {year}
  {1989})}\BibitemShut {NoStop}%
\bibitem [{\citenamefont {Nomura}\ and\ \citenamefont
  {Yamada}(1991)}]{PhysRevB.43.8217}%
  \BibitemOpen
  \bibfield  {author} {\bibinfo {author} {\bibfnamefont {K.}~\bibnamefont
  {Nomura}}\ and\ \bibinfo {author} {\bibfnamefont {M.}~\bibnamefont
  {Yamada}},\ }\href {\doibase 10.1103/PhysRevB.43.8217} {\bibfield  {journal}
  {\bibinfo  {journal} {Phys. Rev. B}\ }\textbf {\bibinfo {volume} {43}},\
  \bibinfo {pages} {8217} (\bibinfo {year} {1991})}\BibitemShut {NoStop}%
\bibitem [{\citenamefont {Eggert}\ \emph {et~al.}(1994)\citenamefont {Eggert},
  \citenamefont {Affleck},\ and\ \citenamefont
  {Takahashi}}]{PhysRevLett.73.332}%
  \BibitemOpen
  \bibfield  {author} {\bibinfo {author} {\bibfnamefont {S.}~\bibnamefont
  {Eggert}}, \bibinfo {author} {\bibfnamefont {I.}~\bibnamefont {Affleck}}, \
  and\ \bibinfo {author} {\bibfnamefont {M.}~\bibnamefont {Takahashi}},\ }\href
  {\doibase 10.1103/PhysRevLett.73.332} {\bibfield  {journal} {\bibinfo
  {journal} {Phys. Rev. Lett.}\ }\textbf {\bibinfo {volume} {73}},\ \bibinfo
  {pages} {332} (\bibinfo {year} {1994})}\BibitemShut {NoStop}%
\bibitem [{\citenamefont {Takigawa}\ \emph {et~al.}(1997)\citenamefont
  {Takigawa}, \citenamefont {Starykh}, \citenamefont {Sandvik},\ and\
  \citenamefont {Singh}}]{PhysRevB.56.13681}%
  \BibitemOpen
  \bibfield  {author} {\bibinfo {author} {\bibfnamefont {M.}~\bibnamefont
  {Takigawa}}, \bibinfo {author} {\bibfnamefont {O.~A.}\ \bibnamefont
  {Starykh}}, \bibinfo {author} {\bibfnamefont {A.~W.}\ \bibnamefont
  {Sandvik}}, \ and\ \bibinfo {author} {\bibfnamefont {R.~R.~P.}\ \bibnamefont
  {Singh}},\ }\href {\doibase 10.1103/PhysRevB.56.13681} {\bibfield  {journal}
  {\bibinfo  {journal} {Phys. Rev. B}\ }\textbf {\bibinfo {volume} {56}},\
  \bibinfo {pages} {13681} (\bibinfo {year} {1997})}\BibitemShut {NoStop}%
\bibitem [{\citenamefont {Affleck}(1998)}]{Affleck_1998}%
  \BibitemOpen
  \bibfield  {author} {\bibinfo {author} {\bibfnamefont {I.}~\bibnamefont
  {Affleck}},\ }\href {\doibase 10.1088/0305-4470/31/20/002} {\bibfield
  {journal} {\bibinfo  {journal} {Journal of Physics A: Mathematical and
  General}\ }\textbf {\bibinfo {volume} {31}},\ \bibinfo {pages} {4573}
  (\bibinfo {year} {1998})}\BibitemShut {NoStop}%
\bibitem [{\citenamefont {Barzykin}(2001)}]{PhysRevB.63.140412}%
  \BibitemOpen
  \bibfield  {author} {\bibinfo {author} {\bibfnamefont {V.}~\bibnamefont
  {Barzykin}},\ }\href {\doibase 10.1103/PhysRevB.63.140412} {\bibfield
  {journal} {\bibinfo  {journal} {Phys. Rev. B}\ }\textbf {\bibinfo {volume}
  {63}},\ \bibinfo {pages} {140412} (\bibinfo {year} {2001})}\BibitemShut
  {NoStop}%
\bibitem [{\citenamefont {Barzykin}(2000)}]{Barzykin_2000}%
  \BibitemOpen
  \bibfield  {author} {\bibinfo {author} {\bibfnamefont {V.}~\bibnamefont
  {Barzykin}},\ }\href {\doibase 10.1088/0953-8984/12/9/309} {\bibfield
  {journal} {\bibinfo  {journal} {Journal of Physics: Condensed Matter}\
  }\textbf {\bibinfo {volume} {12}},\ \bibinfo {pages} {2053} (\bibinfo {year}
  {2000})}\BibitemShut {NoStop}%
\bibitem [{\citenamefont {Dupont}\ \emph {et~al.}(2016)\citenamefont {Dupont},
  \citenamefont {Capponi},\ and\ \citenamefont
  {Laflorencie}}]{PhysRevB.94.144409}%
  \BibitemOpen
  \bibfield  {author} {\bibinfo {author} {\bibfnamefont {M.}~\bibnamefont
  {Dupont}}, \bibinfo {author} {\bibfnamefont {S.}~\bibnamefont {Capponi}}, \
  and\ \bibinfo {author} {\bibfnamefont {N.}~\bibnamefont {Laflorencie}},\
  }\href {\doibase 10.1103/PhysRevB.94.144409} {\bibfield  {journal} {\bibinfo
  {journal} {Phys. Rev. B}\ }\textbf {\bibinfo {volume} {94}},\ \bibinfo
  {pages} {144409} (\bibinfo {year} {2016})}\BibitemShut {NoStop}%
\bibitem [{\citenamefont {Sylju\aa{}sen}\ and\ \citenamefont
  {Sandvik}(2002)}]{syljuaasen2002}%
  \BibitemOpen
  \bibfield  {author} {\bibinfo {author} {\bibfnamefont {O.~F.}\ \bibnamefont
  {Sylju\aa{}sen}}\ and\ \bibinfo {author} {\bibfnamefont {A.~W.}\ \bibnamefont
  {Sandvik}},\ }\href {\doibase 10.1103/PhysRevE.66.046701} {\bibfield
  {journal} {\bibinfo  {journal} {Phys. Rev. E}\ }\textbf {\bibinfo {volume}
  {66}},\ \bibinfo {pages} {046701} (\bibinfo {year} {2002})}\BibitemShut
  {NoStop}%
\bibitem [{\citenamefont {Sandvik}(2010{\natexlab{a}})}]{sandvik2010}%
  \BibitemOpen
  \bibfield  {author} {\bibinfo {author} {\bibfnamefont {A.~W.}\ \bibnamefont
  {Sandvik}},\ }\href {\doibase 10.1063/1.3518900} {\bibfield  {journal}
  {\bibinfo  {journal} {AIP Conf. Proc.}\ }\textbf {\bibinfo {volume} {1297}},\
  \bibinfo {pages} {135} (\bibinfo {year} {2010}{\natexlab{a}})}\BibitemShut
  {NoStop}%
\bibitem [{\citenamefont {Sandvik}(2019)}]{sandvik2019}%
  \BibitemOpen
  \bibfield  {author} {\bibinfo {author} {\bibfnamefont {A.~W.}\ \bibnamefont
  {Sandvik}},\ }\href@noop {} {\emph {\bibinfo {title} {Many-Body Methods for
  Real Materials, Modeling and Simulation}}},\ edited by\ \bibinfo {editor}
  {\bibfnamefont {E.}~\bibnamefont {Pavarini}}, \bibinfo {editor}
  {\bibfnamefont {E.}~\bibnamefont {Koch}}, \ and\ \bibinfo {editor}
  {\bibfnamefont {S.}~\bibnamefont {Zhang}},\ \bibinfo {series} {Verlag des
  Forschungszentrum Julich}, Vol.~\bibinfo {volume} {9}\ (\bibinfo  {publisher}
  {Modeling and Simulation},\ \bibinfo {year} {2019})\BibitemShut {NoStop}%
\bibitem [{\citenamefont {Sandvik}\ and\ \citenamefont
  {Kurkij\"arvi}(1991)}]{PhysRevB.43.5950}%
  \BibitemOpen
  \bibfield  {author} {\bibinfo {author} {\bibfnamefont {A.~W.}\ \bibnamefont
  {Sandvik}}\ and\ \bibinfo {author} {\bibfnamefont {J.}~\bibnamefont
  {Kurkij\"arvi}},\ }\href {\doibase 10.1103/PhysRevB.43.5950} {\bibfield
  {journal} {\bibinfo  {journal} {Phys. Rev. B}\ }\textbf {\bibinfo {volume}
  {43}},\ \bibinfo {pages} {5950} (\bibinfo {year} {1991})}\BibitemShut
  {NoStop}%
\bibitem [{\citenamefont {Fisher}(1964)}]{fisher1964}%
  \BibitemOpen
  \bibfield  {author} {\bibinfo {author} {\bibfnamefont {M.~E.}\ \bibnamefont
  {Fisher}},\ }\href {\doibase 10.1119/1.1970340} {\bibfield  {journal}
  {\bibinfo  {journal} {American Journal of Physics}\ }\textbf {\bibinfo
  {volume} {32}},\ \bibinfo {pages} {343} (\bibinfo {year} {1964})}\BibitemShut
  {NoStop}%
\bibitem [{Note1()}]{Note1}%
  \BibitemOpen
  \bibinfo {note} {In the SU(2) symmetric case, all spin orientations are
  equivalent.}\BibitemShut {Stop}%
\bibitem [{Note2()}]{Note2}%
  \BibitemOpen
  \bibinfo {note} {Note that higher order corrections have also been computed
  analytically~\cite {PhysRevB.63.140412}, $\chi ^{\Delta =1}_{\pi
  }(T)=\protect \frac {\chi '_0}{T}\protect \sqrt {\protect \qopname \relax
  o{ln}(J\Lambda '/T)+\protect \qopname \relax o{ln}\protect \sqrt {\protect
  \qopname \relax o{ln}(J\Lambda '/T)}}$, but our data are best described by
  the simpler form Eq.~\protect \textup {\hbox {\mathsurround \z@ \protect
  \normalfont (\ignorespaces \ref {eq:susc_xxx}\unskip \@@italiccorr
  )}}}\BibitemShut {NoStop}%
\bibitem [{\citenamefont {Starykh}\ \emph {et~al.}(1997)\citenamefont
  {Starykh}, \citenamefont {Sandvik},\ and\ \citenamefont
  {Singh}}]{PhysRevB.55.14953}%
  \BibitemOpen
  \bibfield  {author} {\bibinfo {author} {\bibfnamefont {O.~A.}\ \bibnamefont
  {Starykh}}, \bibinfo {author} {\bibfnamefont {A.~W.}\ \bibnamefont
  {Sandvik}}, \ and\ \bibinfo {author} {\bibfnamefont {R.~R.~P.}\ \bibnamefont
  {Singh}},\ }\href {\doibase 10.1103/PhysRevB.55.14953} {\bibfield  {journal}
  {\bibinfo  {journal} {Phys. Rev. B}\ }\textbf {\bibinfo {volume} {55}},\
  \bibinfo {pages} {14953} (\bibinfo {year} {1997})}\BibitemShut {NoStop}%
\bibitem [{\citenamefont {Kim}\ \emph {et~al.}(1998)\citenamefont {Kim},
  \citenamefont {Greven}, \citenamefont {Wiese},\ and\ \citenamefont
  {Birgeneau}}]{Kim:1998aa}%
  \BibitemOpen
  \bibfield  {author} {\bibinfo {author} {\bibfnamefont {Y.~J.}\ \bibnamefont
  {Kim}}, \bibinfo {author} {\bibfnamefont {M.}~\bibnamefont {Greven}},
  \bibinfo {author} {\bibfnamefont {U.~J.}\ \bibnamefont {Wiese}}, \ and\
  \bibinfo {author} {\bibfnamefont {R.~J.}\ \bibnamefont {Birgeneau}},\ }\href
  {\doibase 10.1007/s100510050382} {\bibfield  {journal} {\bibinfo  {journal}
  {The European Physical Journal B - Condensed Matter and Complex Systems}\
  }\textbf {\bibinfo {volume} {4}},\ \bibinfo {pages} {291} (\bibinfo {year}
  {1998})}\BibitemShut {NoStop}%
\bibitem [{\citenamefont {Xiang}(1998)}]{PhysRevB.58.9142}%
  \BibitemOpen
  \bibfield  {author} {\bibinfo {author} {\bibfnamefont {T.}~\bibnamefont
  {Xiang}},\ }\href {\doibase 10.1103/PhysRevB.58.9142} {\bibfield  {journal}
  {\bibinfo  {journal} {Phys. Rev. B}\ }\textbf {\bibinfo {volume} {58}},\
  \bibinfo {pages} {9142} (\bibinfo {year} {1998})}\BibitemShut {NoStop}%
\bibitem [{\citenamefont {Binder}(1981)}]{PhysRevLett.47.693}%
  \BibitemOpen
  \bibfield  {author} {\bibinfo {author} {\bibfnamefont {K.}~\bibnamefont
  {Binder}},\ }\href {\doibase 10.1103/PhysRevLett.47.693} {\bibfield
  {journal} {\bibinfo  {journal} {Phys. Rev. Lett.}\ }\textbf {\bibinfo
  {volume} {47}},\ \bibinfo {pages} {693} (\bibinfo {year} {1981})}\BibitemShut
  {NoStop}%
\bibitem [{\citenamefont {Botet}\ \emph {et~al.}(1982)\citenamefont {Botet},
  \citenamefont {Jullien},\ and\ \citenamefont {Pfeuty}}]{PhysRevLett.49.478}%
  \BibitemOpen
  \bibfield  {author} {\bibinfo {author} {\bibfnamefont {R.}~\bibnamefont
  {Botet}}, \bibinfo {author} {\bibfnamefont {R.}~\bibnamefont {Jullien}}, \
  and\ \bibinfo {author} {\bibfnamefont {P.}~\bibnamefont {Pfeuty}},\ }\href
  {\doibase 10.1103/PhysRevLett.49.478} {\bibfield  {journal} {\bibinfo
  {journal} {Phys. Rev. Lett.}\ }\textbf {\bibinfo {volume} {49}},\ \bibinfo
  {pages} {478} (\bibinfo {year} {1982})}\BibitemShut {NoStop}%
\bibitem [{\citenamefont {Botet}\ and\ \citenamefont
  {Jullien}(1983)}]{PhysRevB.28.3955}%
  \BibitemOpen
  \bibfield  {author} {\bibinfo {author} {\bibfnamefont {R.}~\bibnamefont
  {Botet}}\ and\ \bibinfo {author} {\bibfnamefont {R.}~\bibnamefont
  {Jullien}},\ }\href {\doibase 10.1103/PhysRevB.28.3955} {\bibfield  {journal}
  {\bibinfo  {journal} {Phys. Rev. B}\ }\textbf {\bibinfo {volume} {28}},\
  \bibinfo {pages} {3955} (\bibinfo {year} {1983})}\BibitemShut {NoStop}%
\bibitem [{\citenamefont {Beach}\ \emph {et~al.}(2005)\citenamefont {Beach},
  \citenamefont {Wang},\ and\ \citenamefont {Sandvik}}]{Beach2005}%
  \BibitemOpen
  \bibfield  {author} {\bibinfo {author} {\bibfnamefont {K.~S.~D.}\
  \bibnamefont {Beach}}, \bibinfo {author} {\bibfnamefont {L.}~\bibnamefont
  {Wang}}, \ and\ \bibinfo {author} {\bibfnamefont {A.~W.}\ \bibnamefont
  {Sandvik}},\ }\href {https://arxiv.org/abs/cond-mat/0505194} {\bibfield
  {journal} {\bibinfo  {journal} {arXiv:cond-mat/0505194}\ } (\bibinfo {year}
  {2005})}\BibitemShut {NoStop}%
\bibitem [{\citenamefont {Wang}\ \emph {et~al.}(2006)\citenamefont {Wang},
  \citenamefont {Beach},\ and\ \citenamefont {Sandvik}}]{wang2006}%
  \BibitemOpen
  \bibfield  {author} {\bibinfo {author} {\bibfnamefont {L.}~\bibnamefont
  {Wang}}, \bibinfo {author} {\bibfnamefont {K.~S.~D.}\ \bibnamefont {Beach}},
  \ and\ \bibinfo {author} {\bibfnamefont {A.~W.}\ \bibnamefont {Sandvik}},\
  }\href {\doibase 10.1103/PhysRevB.73.014431} {\bibfield  {journal} {\bibinfo
  {journal} {Phys. Rev. B}\ }\textbf {\bibinfo {volume} {73}},\ \bibinfo
  {pages} {014431} (\bibinfo {year} {2006})}\BibitemShut {NoStop}%
\bibitem [{Note3()}]{Note3}%
  \BibitemOpen
  \bibinfo {note} {For the SW network, the statistical QMC average result of a
  given simulation is averaged over randomly drawn lattice geometries. Note
  also that because the fitting form is relatively complicated, we have found
  that it can be advantageous to perform, at first, the fit on the scaling
  forms without the corrections to the scaling. We then use the obtained
  fitting parameters as initial guesses for the more complicated form. In all
  cases, we add a small random noise to each initial guess for the
  parameters.}\BibitemShut {Stop}%
\bibitem [{\citenamefont {Corless}\ \emph {et~al.}(1996)\citenamefont
  {Corless}, \citenamefont {Gonnet}, \citenamefont {Hare}, \citenamefont
  {Jeffrey},\ and\ \citenamefont {Knuth}}]{corless_lambertw_1996}%
  \BibitemOpen
  \bibfield  {author} {\bibinfo {author} {\bibfnamefont {R.~M.}\ \bibnamefont
  {Corless}}, \bibinfo {author} {\bibfnamefont {G.~H.}\ \bibnamefont {Gonnet}},
  \bibinfo {author} {\bibfnamefont {D.~E.~G.}\ \bibnamefont {Hare}}, \bibinfo
  {author} {\bibfnamefont {D.~J.}\ \bibnamefont {Jeffrey}}, \ and\ \bibinfo
  {author} {\bibfnamefont {D.~E.}\ \bibnamefont {Knuth}},\ }\href {\doibase
  10.1007/BF02124750} {\bibfield  {journal} {\bibinfo  {journal} {Advances in
  Computational Mathematics}\ }\textbf {\bibinfo {volume} {5}},\ \bibinfo
  {pages} {329} (\bibinfo {year} {1996})}\BibitemShut {NoStop}%
\bibitem [{\citenamefont {Cross}\ and\ \citenamefont
  {Fisher}(1979)}]{PhysRevB.19.402}%
  \BibitemOpen
  \bibfield  {author} {\bibinfo {author} {\bibfnamefont {M.~C.}\ \bibnamefont
  {Cross}}\ and\ \bibinfo {author} {\bibfnamefont {D.~S.}\ \bibnamefont
  {Fisher}},\ }\href {\doibase 10.1103/PhysRevB.19.402} {\bibfield  {journal}
  {\bibinfo  {journal} {Phys. Rev. B}\ }\textbf {\bibinfo {volume} {19}},\
  \bibinfo {pages} {402} (\bibinfo {year} {1979})}\BibitemShut {NoStop}%
\bibitem [{\citenamefont {Sakai}\ and\ \citenamefont
  {Takahashi}(1990)}]{PhysRevB.42.4537}%
  \BibitemOpen
  \bibfield  {author} {\bibinfo {author} {\bibfnamefont {T.}~\bibnamefont
  {Sakai}}\ and\ \bibinfo {author} {\bibfnamefont {M.}~\bibnamefont
  {Takahashi}},\ }\href {\doibase 10.1103/PhysRevB.42.4537} {\bibfield
  {journal} {\bibinfo  {journal} {Phys. Rev. B}\ }\textbf {\bibinfo {volume}
  {42}},\ \bibinfo {pages} {4537} (\bibinfo {year} {1990})}\BibitemShut
  {NoStop}%
\bibitem [{\citenamefont {Wierschem}\ and\ \citenamefont
  {Sengupta}(2014)}]{PhysRevLett.112.247203}%
  \BibitemOpen
  \bibfield  {author} {\bibinfo {author} {\bibfnamefont {K.}~\bibnamefont
  {Wierschem}}\ and\ \bibinfo {author} {\bibfnamefont {P.}~\bibnamefont
  {Sengupta}},\ }\href {\doibase 10.1103/PhysRevLett.112.247203} {\bibfield
  {journal} {\bibinfo  {journal} {Phys. Rev. Lett.}\ }\textbf {\bibinfo
  {volume} {112}},\ \bibinfo {pages} {247203} (\bibinfo {year}
  {2014})}\BibitemShut {NoStop}%
\bibitem [{\citenamefont {Klanj\ifmmode~\check{s}\else \v{s}\fi{}ek}\ \emph
  {et~al.}(2008)\citenamefont {Klanj\ifmmode~\check{s}\else \v{s}\fi{}ek},
  \citenamefont {Mayaffre}, \citenamefont {Berthier}, \citenamefont
  {Horvati\ifmmode~\acute{c}\else \'{c}\fi{}}, \citenamefont {Chiari},
  \citenamefont {Piovesana}, \citenamefont {Bouillot}, \citenamefont {Kollath},
  \citenamefont {Orignac}, \citenamefont {Citro},\ and\ \citenamefont
  {Giamarchi}}]{PhysRevLett.101.137207}%
  \BibitemOpen
  \bibfield  {author} {\bibinfo {author} {\bibfnamefont {M.}~\bibnamefont
  {Klanj\ifmmode~\check{s}\else \v{s}\fi{}ek}}, \bibinfo {author}
  {\bibfnamefont {H.}~\bibnamefont {Mayaffre}}, \bibinfo {author}
  {\bibfnamefont {C.}~\bibnamefont {Berthier}}, \bibinfo {author}
  {\bibfnamefont {M.}~\bibnamefont {Horvati\ifmmode~\acute{c}\else
  \'{c}\fi{}}}, \bibinfo {author} {\bibfnamefont {B.}~\bibnamefont {Chiari}},
  \bibinfo {author} {\bibfnamefont {O.}~\bibnamefont {Piovesana}}, \bibinfo
  {author} {\bibfnamefont {P.}~\bibnamefont {Bouillot}}, \bibinfo {author}
  {\bibfnamefont {C.}~\bibnamefont {Kollath}}, \bibinfo {author} {\bibfnamefont
  {E.}~\bibnamefont {Orignac}}, \bibinfo {author} {\bibfnamefont
  {R.}~\bibnamefont {Citro}}, \ and\ \bibinfo {author} {\bibfnamefont
  {T.}~\bibnamefont {Giamarchi}},\ }\href {\doibase
  10.1103/PhysRevLett.101.137207} {\bibfield  {journal} {\bibinfo  {journal}
  {Phys. Rev. Lett.}\ }\textbf {\bibinfo {volume} {101}},\ \bibinfo {pages}
  {137207} (\bibinfo {year} {2008})}\BibitemShut {NoStop}%
\bibitem [{\citenamefont {Furuya}\ \emph {et~al.}(2016)\citenamefont {Furuya},
  \citenamefont {Dupont}, \citenamefont {Capponi}, \citenamefont
  {Laflorencie},\ and\ \citenamefont {Giamarchi}}]{PhysRevB.94.144403}%
  \BibitemOpen
  \bibfield  {author} {\bibinfo {author} {\bibfnamefont {S.~C.}\ \bibnamefont
  {Furuya}}, \bibinfo {author} {\bibfnamefont {M.}~\bibnamefont {Dupont}},
  \bibinfo {author} {\bibfnamefont {S.}~\bibnamefont {Capponi}}, \bibinfo
  {author} {\bibfnamefont {N.}~\bibnamefont {Laflorencie}}, \ and\ \bibinfo
  {author} {\bibfnamefont {T.}~\bibnamefont {Giamarchi}},\ }\href {\doibase
  10.1103/PhysRevB.94.144403} {\bibfield  {journal} {\bibinfo  {journal} {Phys.
  Rev. B}\ }\textbf {\bibinfo {volume} {94}},\ \bibinfo {pages} {144403}
  (\bibinfo {year} {2016})}\BibitemShut {NoStop}%
\bibitem [{\citenamefont {Giraud}\ \emph {et~al.}(2005)\citenamefont {Giraud},
  \citenamefont {Georgeot},\ and\ \citenamefont
  {Shepelyansky}}]{PhysRevE.72.036203}%
  \BibitemOpen
  \bibfield  {author} {\bibinfo {author} {\bibfnamefont {O.}~\bibnamefont
  {Giraud}}, \bibinfo {author} {\bibfnamefont {B.}~\bibnamefont {Georgeot}}, \
  and\ \bibinfo {author} {\bibfnamefont {D.~L.}\ \bibnamefont {Shepelyansky}},\
  }\href {\doibase 10.1103/PhysRevE.72.036203} {\bibfield  {journal} {\bibinfo
  {journal} {Phys. Rev. E}\ }\textbf {\bibinfo {volume} {72}},\ \bibinfo
  {pages} {036203} (\bibinfo {year} {2005})}\BibitemShut {NoStop}%
\bibitem [{\citenamefont {Garc\'{\i}a-Mata}\ \emph {et~al.}(2017)\citenamefont
  {Garc\'{\i}a-Mata}, \citenamefont {Giraud}, \citenamefont {Georgeot},
  \citenamefont {Martin}, \citenamefont {Dubertrand},\ and\ \citenamefont
  {Lemari\'e}}]{PhysRevLett.118.166801}%
  \BibitemOpen
  \bibfield  {author} {\bibinfo {author} {\bibfnamefont {I.}~\bibnamefont
  {Garc\'{\i}a-Mata}}, \bibinfo {author} {\bibfnamefont {O.}~\bibnamefont
  {Giraud}}, \bibinfo {author} {\bibfnamefont {B.}~\bibnamefont {Georgeot}},
  \bibinfo {author} {\bibfnamefont {J.}~\bibnamefont {Martin}}, \bibinfo
  {author} {\bibfnamefont {R.}~\bibnamefont {Dubertrand}}, \ and\ \bibinfo
  {author} {\bibfnamefont {G.}~\bibnamefont {Lemari\'e}},\ }\href {\doibase
  10.1103/PhysRevLett.118.166801} {\bibfield  {journal} {\bibinfo  {journal}
  {Phys. Rev. Lett.}\ }\textbf {\bibinfo {volume} {118}},\ \bibinfo {pages}
  {166801} (\bibinfo {year} {2017})}\BibitemShut {NoStop}%
\bibitem [{\citenamefont {Garc\'{\i}a-Mata}\ \emph {et~al.}(2020)\citenamefont
  {Garc\'{\i}a-Mata}, \citenamefont {Martin}, \citenamefont {Dubertrand},
  \citenamefont {Giraud}, \citenamefont {Georgeot},\ and\ \citenamefont
  {Lemari\'e}}]{PhysRevResearch.2.012020}%
  \BibitemOpen
  \bibfield  {author} {\bibinfo {author} {\bibfnamefont {I.}~\bibnamefont
  {Garc\'{\i}a-Mata}}, \bibinfo {author} {\bibfnamefont {J.}~\bibnamefont
  {Martin}}, \bibinfo {author} {\bibfnamefont {R.}~\bibnamefont {Dubertrand}},
  \bibinfo {author} {\bibfnamefont {O.}~\bibnamefont {Giraud}}, \bibinfo
  {author} {\bibfnamefont {B.}~\bibnamefont {Georgeot}}, \ and\ \bibinfo
  {author} {\bibfnamefont {G.}~\bibnamefont {Lemari\'e}},\ }\href {\doibase
  10.1103/PhysRevResearch.2.012020} {\bibfield  {journal} {\bibinfo  {journal}
  {Phys. Rev. Research}\ }\textbf {\bibinfo {volume} {2}},\ \bibinfo {pages}
  {012020} (\bibinfo {year} {2020})}\BibitemShut {NoStop}%
\bibitem [{\citenamefont {Dupont}\ \emph {et~al.}(2020)\citenamefont {Dupont},
  \citenamefont {Laflorencie},\ and\ \citenamefont
  {Lemari\'e}}]{PhysRevB.102.174205}%
  \BibitemOpen
  \bibfield  {author} {\bibinfo {author} {\bibfnamefont {M.}~\bibnamefont
  {Dupont}}, \bibinfo {author} {\bibfnamefont {N.}~\bibnamefont {Laflorencie}},
  \ and\ \bibinfo {author} {\bibfnamefont {G.}~\bibnamefont {Lemari\'e}},\
  }\href {\doibase 10.1103/PhysRevB.102.174205} {\bibfield  {journal} {\bibinfo
   {journal} {Phys. Rev. B}\ }\textbf {\bibinfo {volume} {102}},\ \bibinfo
  {pages} {174205} (\bibinfo {year} {2020})}\BibitemShut {NoStop}%
\bibitem [{\citenamefont {Fisher}(1995)}]{PhysRevB.51.6411}%
  \BibitemOpen
  \bibfield  {author} {\bibinfo {author} {\bibfnamefont {D.~S.}\ \bibnamefont
  {Fisher}},\ }\href {\doibase 10.1103/PhysRevB.51.6411} {\bibfield  {journal}
  {\bibinfo  {journal} {Phys. Rev. B}\ }\textbf {\bibinfo {volume} {51}},\
  \bibinfo {pages} {6411} (\bibinfo {year} {1995})}\BibitemShut {NoStop}%
\bibitem [{\citenamefont {Carpentier}\ \emph {et~al.}(2005)\citenamefont
  {Carpentier}, \citenamefont {Pujol},\ and\ \citenamefont
  {Giering}}]{PhysRevE.72.066101}%
  \BibitemOpen
  \bibfield  {author} {\bibinfo {author} {\bibfnamefont {D.}~\bibnamefont
  {Carpentier}}, \bibinfo {author} {\bibfnamefont {P.}~\bibnamefont {Pujol}}, \
  and\ \bibinfo {author} {\bibfnamefont {K.-U.}\ \bibnamefont {Giering}},\
  }\href {\doibase 10.1103/PhysRevE.72.066101} {\bibfield  {journal} {\bibinfo
  {journal} {Phys. Rev. E}\ }\textbf {\bibinfo {volume} {72}},\ \bibinfo
  {pages} {066101} (\bibinfo {year} {2005})}\BibitemShut {NoStop}%
\bibitem [{\citenamefont
  {Sandvik}(2010{\natexlab{b}})}]{PhysRevLett.104.137204}%
  \BibitemOpen
  \bibfield  {author} {\bibinfo {author} {\bibfnamefont {A.~W.}\ \bibnamefont
  {Sandvik}},\ }\href {\doibase 10.1103/PhysRevLett.104.137204} {\bibfield
  {journal} {\bibinfo  {journal} {Phys. Rev. Lett.}\ }\textbf {\bibinfo
  {volume} {104}},\ \bibinfo {pages} {137204} (\bibinfo {year}
  {2010}{\natexlab{b}})}\BibitemShut {NoStop}%
\bibitem [{\citenamefont {Bentsen}\ \emph
  {et~al.}(2019{\natexlab{a}})\citenamefont {Bentsen}, \citenamefont
  {Potirniche}, \citenamefont {Bulchandani}, \citenamefont {Scaffidi},
  \citenamefont {Cao}, \citenamefont {Qi}, \citenamefont {Schleier-Smith},\
  and\ \citenamefont {Altman}}]{PhysRevX.9.041011}%
  \BibitemOpen
  \bibfield  {author} {\bibinfo {author} {\bibfnamefont {G.}~\bibnamefont
  {Bentsen}}, \bibinfo {author} {\bibfnamefont {I.-D.}\ \bibnamefont
  {Potirniche}}, \bibinfo {author} {\bibfnamefont {V.~B.}\ \bibnamefont
  {Bulchandani}}, \bibinfo {author} {\bibfnamefont {T.}~\bibnamefont
  {Scaffidi}}, \bibinfo {author} {\bibfnamefont {X.}~\bibnamefont {Cao}},
  \bibinfo {author} {\bibfnamefont {X.-L.}\ \bibnamefont {Qi}}, \bibinfo
  {author} {\bibfnamefont {M.}~\bibnamefont {Schleier-Smith}}, \ and\ \bibinfo
  {author} {\bibfnamefont {E.}~\bibnamefont {Altman}},\ }\href {\doibase
  10.1103/PhysRevX.9.041011} {\bibfield  {journal} {\bibinfo  {journal} {Phys.
  Rev. X}\ }\textbf {\bibinfo {volume} {9}},\ \bibinfo {pages} {041011}
  (\bibinfo {year} {2019}{\natexlab{a}})}\BibitemShut {NoStop}%
\bibitem [{\citenamefont {Bentsen}\ \emph
  {et~al.}(2019{\natexlab{b}})\citenamefont {Bentsen}, \citenamefont
  {Hashizume}, \citenamefont {Buyskikh}, \citenamefont {Davis}, \citenamefont
  {Daley}, \citenamefont {Gubser},\ and\ \citenamefont
  {Schleier-Smith}}]{PhysRevLett.123.130601}%
  \BibitemOpen
  \bibfield  {author} {\bibinfo {author} {\bibfnamefont {G.}~\bibnamefont
  {Bentsen}}, \bibinfo {author} {\bibfnamefont {T.}~\bibnamefont {Hashizume}},
  \bibinfo {author} {\bibfnamefont {A.~S.}\ \bibnamefont {Buyskikh}}, \bibinfo
  {author} {\bibfnamefont {E.~J.}\ \bibnamefont {Davis}}, \bibinfo {author}
  {\bibfnamefont {A.~J.}\ \bibnamefont {Daley}}, \bibinfo {author}
  {\bibfnamefont {S.~S.}\ \bibnamefont {Gubser}}, \ and\ \bibinfo {author}
  {\bibfnamefont {M.}~\bibnamefont {Schleier-Smith}},\ }\href {\doibase
  10.1103/PhysRevLett.123.130601} {\bibfield  {journal} {\bibinfo  {journal}
  {Phys. Rev. Lett.}\ }\textbf {\bibinfo {volume} {123}},\ \bibinfo {pages}
  {130601} (\bibinfo {year} {2019}{\natexlab{b}})}\BibitemShut {NoStop}%
\bibitem [{\citenamefont {Davis}\ \emph {et~al.}(2020)\citenamefont {Davis},
  \citenamefont {Periwal}, \citenamefont {Cooper}, \citenamefont {Bentsen},
  \citenamefont {Evered}, \citenamefont {Van~Kirk},\ and\ \citenamefont
  {Schleier-Smith}}]{PhysRevLett.125.060402}%
  \BibitemOpen
  \bibfield  {author} {\bibinfo {author} {\bibfnamefont {E.~J.}\ \bibnamefont
  {Davis}}, \bibinfo {author} {\bibfnamefont {A.}~\bibnamefont {Periwal}},
  \bibinfo {author} {\bibfnamefont {E.~S.}\ \bibnamefont {Cooper}}, \bibinfo
  {author} {\bibfnamefont {G.}~\bibnamefont {Bentsen}}, \bibinfo {author}
  {\bibfnamefont {S.~J.}\ \bibnamefont {Evered}}, \bibinfo {author}
  {\bibfnamefont {K.}~\bibnamefont {Van~Kirk}}, \ and\ \bibinfo {author}
  {\bibfnamefont {M.~H.}\ \bibnamefont {Schleier-Smith}},\ }\href {\doibase
  10.1103/PhysRevLett.125.060402} {\bibfield  {journal} {\bibinfo  {journal}
  {Phys. Rev. Lett.}\ }\textbf {\bibinfo {volume} {125}},\ \bibinfo {pages}
  {060402} (\bibinfo {year} {2020})}\BibitemShut {NoStop}%
\end{thebibliography}%

\end{document}